\documentclass[%
 reprint,
 amsmath,amssymb,
 aps,
]{revtex4-2}

\usepackage{subfigure}
\usepackage{graphicx}% Include figure files
\usepackage{dcolumn}% Align table columns on decimal point
\usepackage{bm}% bold math
\usepackage{hyperref}% add hypertext capabilities
\DeclareMathOperator{\tr}{tr}

\begin{document}

\preprint{APS/123-QED}

\title{Quantum entanglement, local indicators and effect of external fields in the Kugel-Khomskii model
} % Force line breaks with \\

\author{V.~E.~Valiulin}
\email{valiulin@phystech.edu}
\affiliation{Moscow Institute of Physics and Technology (National Research University), Dolgoprudny 141701, Russia}
\affiliation{Institute for High Pressure Physics, Russian Academy of Sciences, Moscow (Troitsk) 108840, Russia}

\author{A.~V.~Mikheyenkov}
\affiliation{Moscow Institute of Physics and Technology (National Research University), Dolgoprudny 141701, Russia}
\affiliation{Institute for High Pressure Physics, Russian Academy of Sciences, Moscow (Troitsk) 108840, Russia}

\author{N.~M.~Chtchelkatchev}
\affiliation{Moscow Institute of Physics and Technology (National Research University), Dolgoprudny 141701, Russia}
\affiliation{Institute for High Pressure Physics, Russian Academy of Sciences, Moscow (Troitsk) 108840, Russia}

\author{K.~I.~Kugel}
\affiliation{Institute for Theoretical and
Applied Electrodynamics, Russian Academy of Sciences, Moscow 125412, Russia}
\affiliation{National Research University Higher School of Economics, Moscow 101000, Russia}

\date{\today}% It is always \today, today,

\begin{abstract}
Using the exact diagonalization technique, we determine the energy spectrum and wave functions for finite chains described by the two-spin (Kugel--Khomskii) model with different types of intersubsystem exchange terms. The found solutions provide a possibility to address the problem of quantum entanglement inherent to this class of models. We put the main emphasis on the calculations of the concurrence treated as an adequate numerical measure of the entanglement. We also analyze the behavior of two-site correlation functions considered as a local indicator of entanglement. We construct the phase diagrams of the models involving the regions of nonzero entanglement. The pronounced effect of external fields, conjugated to both spin variables on the regions with entanglement, could both enhance and weaken the entanglement depending on the parameters of the models.
\end{abstract}

\maketitle

%\tableofcontents

\section{\label{sec:intro} Introduction}

Entanglement is one of the main manifestations of the quantum nature of the matter being intensively studied in connection with the development of quantum computers~\cite{Vidal03_PRL,JozsaPRSA2003,Bengts06_Book,Horode09_RMP,
CuffaroPhilSci2013,ZidanAppMathInfSci2018,Chen20_PRL,Gale20_PRA,George20_NRP,Grimsm20_PRX,McArdl20_RMP,Rogger20_PRD,Wu20_PRR}
The problem of entanglement has been studied in detail for nanosystems, especially for quantum dots~\cite{JosefssonPRB2020}. Such systems are used for the design of quantum information processing systems.

In the solids, traditional for condensed matter physics, things are not so clear. Many standard solid-state systems are entangled. There can be no doubt, that electrons in a metal are entangled~\cite{BeenakkerLect2006}, but how is to verify this directly experimentally, using the accepted criteria. The conventional method for determining entanglement (we do not mention here Bell inequalities~\cite{ChtchelkatchevPRB2002,HyllusPRA2005}, very efficient in optics, but not in solids, and other exotic methods) involves the determination of the density matrix, which is quite computationally problematic, even for a relatively small cluster. The most promising would be to extract information about entanglement from correlation functions related directly to the system in hand. There are many efficient methods for calculating correlation functions, both numerically and analytically for strongly correlated systems with a large number of degrees of freedom, in particular, in the thermodynamic limit. Moreover, many correlators are experimentally determined. Another important issue is how one can manage the degree of entanglement. The influence of external fields on entanglement is crucial here, since it provides a possibility to control an entangled system in quantum information processing.
We note, that these fields may have a completely different nature, from the magnetic field to elastic stresses.

The most vivid example of the entanglement in condensed matter is represented by the models  involving two kinds of interacting spin variables. Two-spin models themselves usually appear in the description of specific features of transition metal compounds with the coupled spin and orbital degrees of freedom; that is why such models are often referred to as spin--orbital ones (sometimes, the term Kugel--Khomskii model is used)~\cite{Kugel82_SPU,TokuraScience2000,Oles12_JPCM}. Unusual effects related to the spin--orbital correlations and the corresponding quantum entanglement are widely discussed in the current literature. In particular, the possibility of extraordinary spin--orbital quantum states and transitions between them was pointed out~\cite{Brzezi13_PRB,Belemu17_PRB,Belemu18_NJP,Valiul19_JL,Gotfry20_PRR}.
~\cite{Baldin20_NP,Fumaga20_PRB}.

The simplest version of the Kugel-Khomskii model --- the $SU(2) \times SU(2)$ model with $SU(2)$ symmetries for both spin-$1/2$ and pseudospin-$1/2$ operators ($\hat{\mathbf{S}}$ and $\hat{\mathbf{T}}$) and a positive factor at spin-pseudospin interaction was used in an early attempt in the context of the entanglement~\cite{Chen07_PRB}.

Later on, the entanglement was sought in various other related models: of $SU(2) \times XY$ \cite{Brzezi14_PRLa}, $SU(2) \times XXZ$~\cite{You15_PRB}, and $SU(2) \times SU(2)$ with additional spin-orbit anisotropy~\cite{Gotfry20_PRR}.
Briefly, the results of this analysis come to detection and characterization of the significant entanglement area,
the degree of the entanglement (mainly through the von Neumann entropy), and sometimes indication of possible complex entangled excitations
\cite{You12_PRB}. All the mentioned works estimate the entanglement, the phase boundaries etc. numerically for finite chains.

In contrast to the cited works, here we focus on how to manage the degree of entanglement. This can be done by mixing different intrasubsystem and intersubsystem interactions and by applying and switching external fields. Here we consider these two issues.

We consider several versions of spin-orbital model both with symmetric and nonsymmetric spin-pseudospin interaction. We also introduce different kinds of external fields and study their effect on the entanglement.
In addition, we show the relationship between the degree of entanglement and pair correlators between the orbital and spin degrees of freedom.

In general, the Hamiltonian of the model reads
\begin{equation}\label{eqKHH}
\widehat{\bf{H}} = \widehat{\bf{H}}_{s} + \widehat{\bf{H}}_{t} + \widehat{\bf{H}}_{ts},
\end{equation}
Here $\widehat{\bf{H}}_{s}$, $\widehat{\bf{H}}_{t}$ are Heisenberg-type interactions in the spin pseudospin-spin subsystems:
\begin{equation}\label{eqKHS}
\widehat{\bf{H}}_{s} = J\sum_{<\bf{i},\bf{j}>}
{\widehat{\bf{S}}}_{\bf{i}}{\widehat{\bf{S}}}_{\bf{j}};\qquad \widehat{\bf{H}}_{t} = I\sum_{<\bf{i},\bf{j}>} {\widehat{\bf{T}}}_{\bf{i}}{\widehat{\bf{T}}}_{\bf{j}},
\end{equation}
and $\widehat{\bf{H}}_{ts}$ is interaction between subsystems. Depending on the compound and its symmetry $\widehat{\bf{H}}_{ts}$ could be written as:
\begin{gather}
\widehat{\bf{H}}_{ts}^{(1)} = K \sum_{<\mathbf{{i},{j}>}}\left( {\widehat{\mathbf{S}}}_{\mathbf{i}}{\widehat{\mathbf{S}}}_{\mathbf{j}}\right) \left( {\widehat{\mathbf{T}}}_{\mathbf{i}}{\widehat{\mathbf{T}}}_{\mathbf{j}}\right),
\label{eqKHTS1}
\\
\widehat{\bf{H}}_{ts}^{(2)} = K \sum_{<\mathbf{{i},{j}>}}\left( {\widehat{\mathbf{S}}}_{\mathbf{i}}{\widehat{\mathbf{S}}}_{\mathbf{j}}\right) \left( T_{\mathbf{i}}^{z}T_{\mathbf{j}}^{z}\right),
\label{eqKHTS2}
\\
\widehat{\bf{H}}_{ts}^{(3)} = K \sum_{<\mathbf{{i},{j}>}}\left( S_{\mathbf{i}}^{z}S_{\mathbf{j}}^{z}\right) \left( T_{\mathbf{i}}^{z}T_{\mathbf{j}}^{z}\right),
\label{eqKHTS3}
\\
\widehat{\bf{H}}_{ts}^{(4)} = K \sum_{<\mathbf{{i},{j}>}\alpha }\left( S_{\mathbf{i}}^{\alpha }S_{\mathbf{j}}^{\alpha }T_{\mathbf{i}}^{\alpha }T_{\mathbf{j}}^{\alpha }\right),
\label{eqKHTS4}
\end{gather}
In (\ref{eqKHS})--(\ref{eqKHTS4}) $\bf{i},\bf{j}$ are vectors of the nearest neighbors, $\widehat{\bf{S}}_{\bf{i}}$ and $\widehat{\bf{T}}_{\bf{i}}$ are spin and pseudospain operators, related to orbital degrees of freedom. Hereafter, we consider the most common case when $S=1/2$, $T=1/2$. $\alpha$ is a spin and pseudospin components index.

Note here that a broad class of Hamiltonians of this type can be simulated not only in the framework of solid-state strongly correlated systems but also by ultracold atoms in the traps~\cite{Belemu17_PRB, Belemu18_NJP}.
In this case, the Kugel-Khomskii model may be applicable to the bosonic atoms with an integer spin. Note also that in transition metal compounds (such as ruthenates or vanadates), we are sometimes dealing with integer values of effective spin and orbital quantum numbers.

The additional terms to the Hamiltonian related to the presence of external magnetic fields in both subsystems can be written as:
\begin{equation}\label{eqKHf}
\widehat{\bf{H}}_{f} = -\mathcal{H}_{s}\sum_{\bf{i}}
{\widehat{\bf{S}}}_{\bf{i}}^{z} - \mathcal{H}_{t}\sum_{\bf{i}}
{\widehat{\bf{T}}}_{\bf{i}}^{z},
\end{equation}
where $\mathcal{H}_{s}$, $\mathcal{H}_{t}$ fields in spin and pseudospin systems, respectively. An efficient magnetic field in a pseudospin system occurs, for example, as a result of the action of elastic stresses during uniaxial compression of a crystal. We note that in this model, in contrast to multi-sublattice magnets, the fields $\mathcal{H}_{s}$, $\mathcal{H}_{t}$ can be steered in opposite directions.
Moreover, hereafter we consider also staggered fields in both subsystems.

The entanglement of the two systems can be determined if density matrix is known. There are several quantitative criteria divided into two main courses. One is based on the calculation of von Neumann entropy \cite{Chen07_PRB,You15_PRB,Gotfry20_PRR,You12_PRB,Oles12_JPCM}, while the second one requires a partial trace of the density matrix by the degrees of freedom of one of the subsystems. We note right away that qualitatively all criteria give the same result. Nonetheless, they may differ quantitatively. Here, we use the so called ``concurrence''. Naturally, since we use the exact diagonalization of the Hamiltonian
\cite{Luesche09_PRB,Medved17_PRB,Schiff19_PRA,Wang19_PRB,Tanaka19_PRB} method, any other criterion can also be calculated.

As it was mentioned, we study entanglement, between two subsystems --- spin and orbital. Concurrence \cite{Horode09_RMP} is defined as
\begin{equation}\label{conc}
C = \sqrt{2(1-\tr_{1}(\tr_{2}(\widehat{\rho})^{2}))},
\end{equation}
where $\widehat{\rho}$ is the density matrix of the entire system,  $\tr_{i}(\widehat{\rho}))$ is the partial trace of the density matrix in one of the subsystems, $i$ is the subsystem index (in our case, spin or pseudospin). Thus defined concurrence for two single particles takes values from $C = 0$ in the absence of entanglement, to $C = \sqrt{3/2}$ in the textbook Einstein-Podolsky-Rosen pair.

We compare the entanglement obtained in terms of the strict criterion based on $C$ with the behavior of the local correlation functions of the operators $\hat{\mathbf{S}}_{\mathbf{i}}$ and $\hat{\mathbf{T}}_{\mathbf{j}}$. It turns out that paired correlators provide minimal information about entanglement, even if the operators belong to different cites. Moreover, the range of parameters where the state of the system is most entangled could be found with the correlators of the four operators, more precisely, their gradients.

Naturally, the inclusion of sufficiently high uniform external magnetic fields~\eqref{eqKHf} suppresses entanglement. Nonetheless, in the range of interest, when the magnetic field has the same order of amplitude as the exchange integrals $J,I,K$, entanglement is not suppressed. Furthermore, as it will be seen below, in some cases the external field surprisingly increases the entanglement.
There is a dramatic change, however, in the regions with the strongest entanglement in the phase diagram. The most vivid effect is the shrink of entanglement areas along specific directions or at points in the phase diagram under the influence of external fields.

\section{\label{sec:methods} Methods}

We consider the Kugel-Khomskii model (\ref{eqKHH})--(\ref{eqKHS}) with the conventional  symmetric spin-pseudospin interaction (\ref{eqKHTS1}) and the related models with asymmetric (\ref{eqKHTS2}), (\ref{eqKHTS3}) and symmetric (\ref{eqKHTS4}) interactions for a small linear cluster. We accurately determine the many-particle ground state wave function in the framework of the exact diagonalization method.
The maximum cluster size is limited by computing resources, nevertheless the key characteristics of the system are stable for variations in the chain size. We study both the cases of zero field and strong external field in each subsystem, and focus mainly on how to
manage the degree of entanglement.

This leads to a nontrivial and unobvious behaviour of entanglement between spin and orbital degrees of freedom.

We have studied in detail one-dimensional systems with different boundary conditions: an open chain and a ring. For the whole range of the considered parameters, the open chain appeared to be more convenient for calculation.
In addition, as it was mentioned earlier, we consider mainly the case of nonzero external fields when the problem with the ground state accidental degeneracy is insignificant (for zero field limit we simply set relatively small fields). We should also note that the anisotropy~\cite{Gotfry20_PRR} removes the problem even without external fields.

Hereinafter, we consider the open chain by the exact diagonalization method. We calculate the ground state wave function, that allows us to evaluate von Neumann entropy, any entanglement criterion, as well as correlation functions in each subsystem and between them. The Hamiltonian matrices for the systems under study are very sparse, so it is natural to use the sparse matrix format.
The maximum available size of the chain for comprehensive calculation is determined by the computational resources, mainly by the RAM size, so we extrapolate the results to $1/N \to 0$.

In our work, we have mainly used the QuTiP package, which simplifies the work with quantum objects~\cite{Johans12_CPC,Johans13_CPC}.
In particular, the package has a  convenient interface for constructing the many-particle Hamiltonian using a large number of direct products of various spin operators. All objects in the package are by default converted to sparse format, which significantly simplifies their further processing. The exact diagonalization procedure was performed in the QuTiP package as well. A typical calculation for a chain of 10 cites for 3600 points takes about a day.
Results for $N = 8,9,10$ slightly differ qualitatively and allow fine extrapolation to $1/N \to 0$. When it was possible, we have compared the results with the earlier works on entanglement.

\section{\label{sec:sstt} Kugel-Khomskii model with
$\hat{\bf{H}}_{ts} = \sum ({\hat{\mathbf{S}}}_{\mathbf{i}}{\hat{\mathbf{S}}}_{\mathbf{j}}) ( {\hat{\mathbf{T}}}_{\mathbf{i}}{\hat{\mathbf{T}}}_{\mathbf{j}})$
interaction}

First, we consider the Kugel-Khomskii model (\ref{eqKHH})--(\ref{eqKHS}) with the most common form (\ref{eqKHTS1}) of spin-pseudospin interaction.
We remind that in the mean field, all four common phases are realized: FM-FM, AFM-AFM, FM-AFM, and AFM-FM~\cite{Brink98_PRB}. For large absolute values of $K<0$ compared to $I$ and $J$, this system prefers FM or AFM ordering in both subsystems simultaneously. The opposite case, large $K>0$, favors FM in one subsystem and AFM in the other.

For infinite system, quantum fluctuations destroy long-range order even at $T \to 0$ and the state structure is governed by the local order,
i.e. correlation functions on distinct sites. We address a finite chain, but not to go into the redundant details, will mark different phases (technically, different local orders) by local correlation functions.

In the mean field, FM order in e.g. spin subsystem can be characterized by unidirectional average of spins
$\langle {\hat{\mathbf{S}}}_{\mathbf{i}} \rangle$, and AFM order ---
by a checkerboard pattern (in 1D average spin directions altering from site to site). In terms of local correlators (irrespective to the long-range order)
FM structure corresponds to
$\langle {\hat{\mathbf{S}}}_{\mathbf{i}}{\hat{\mathbf{S}}}_{\mathbf{j}}\rangle > 0$ for any pair of sites.
As for AFM, the sign of correlation function $\langle {\hat{\mathbf{S}}}_{\mathbf{i}}{\hat{\mathbf{S}}}_{\mathbf{j}}\rangle$ is negative for nearest neighbor sites $\mathbf{i}$, $\mathbf{j}$, and alters when $\mathbf{i}$ and $\mathbf{j}$ make a step one from the another.
The same naturally holds for the pseudospin subsystem.

In the quantum case, we adopt the following classification:
``FM'' --- ${\langle \hat{\mathbf{S}}}_{\mathbf{i}} {\hat{\mathbf{S}}}_{\mathbf{j}}\rangle > 0$
for close neighbor pairs $\mathbf{i}$, $\mathbf{j}$;
``AFM'' --- ${\langle \hat{\mathbf{S}}}_{\mathbf{i}} {\hat{\mathbf{S}}}_{\mathbf{j}}\rangle < 0$ for nearest neighbors and altering henceforth.
We do not deal with the exhaustive classification and the fine details of the state structure, but rather superficially mark the local correlation.
All the foregoing does not necessarily mean the phase transitions with distinct order parameters, but rather short-range order rearrangement.

Hereafter, we study how entanglement changes across local order boundaries, i.e. among the areas with different patterns
of local correlations.

\subsection{Entanglement and the sign of intersubsystem exchange $K$}

We begin with the case of a negative intersubsystem exchange $K<0$.

\begin{figure}[t]
(a) \includegraphics[width=0.60\columnwidth]{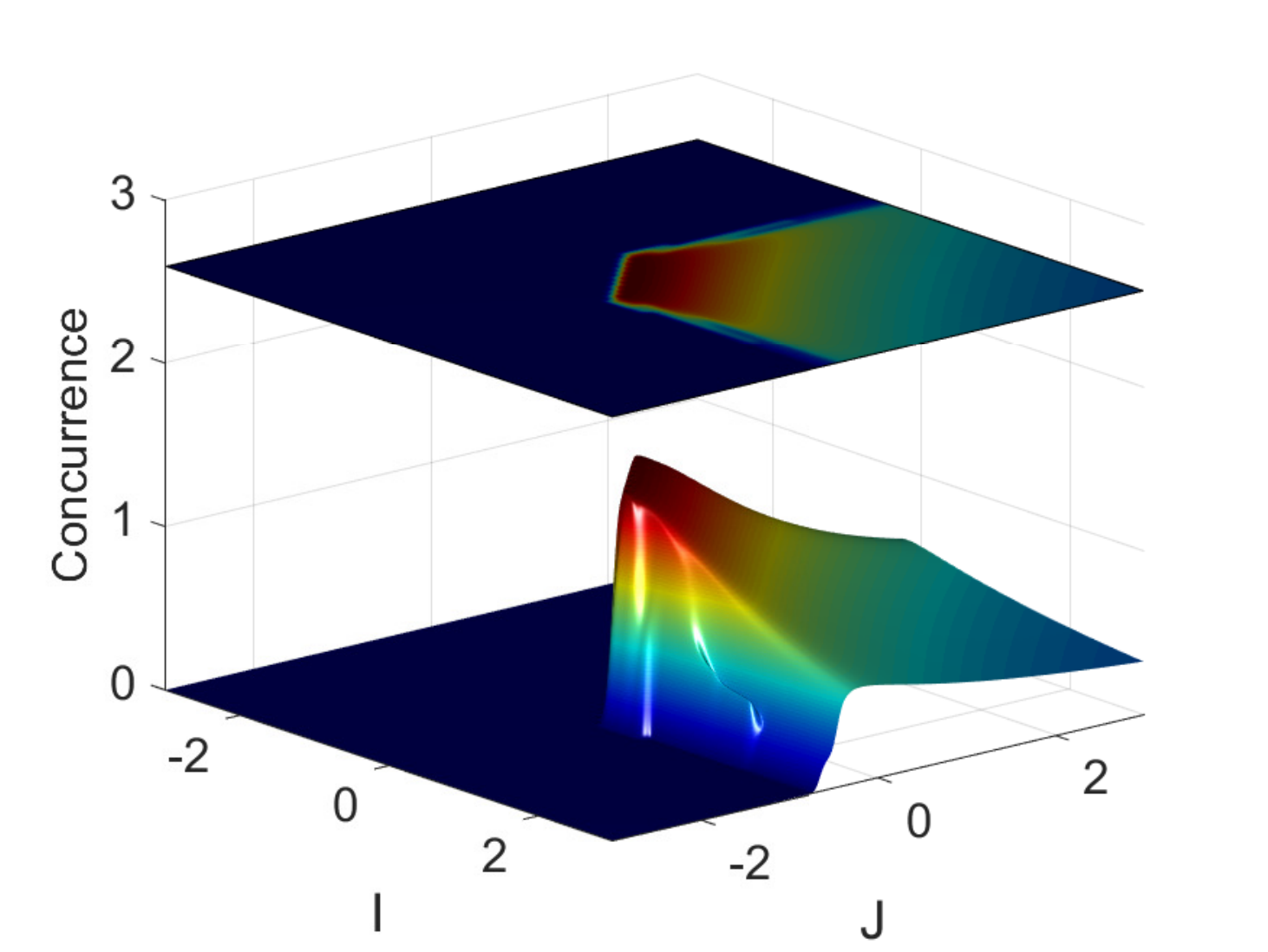}\\
(b) \includegraphics[width=0.43\columnwidth]{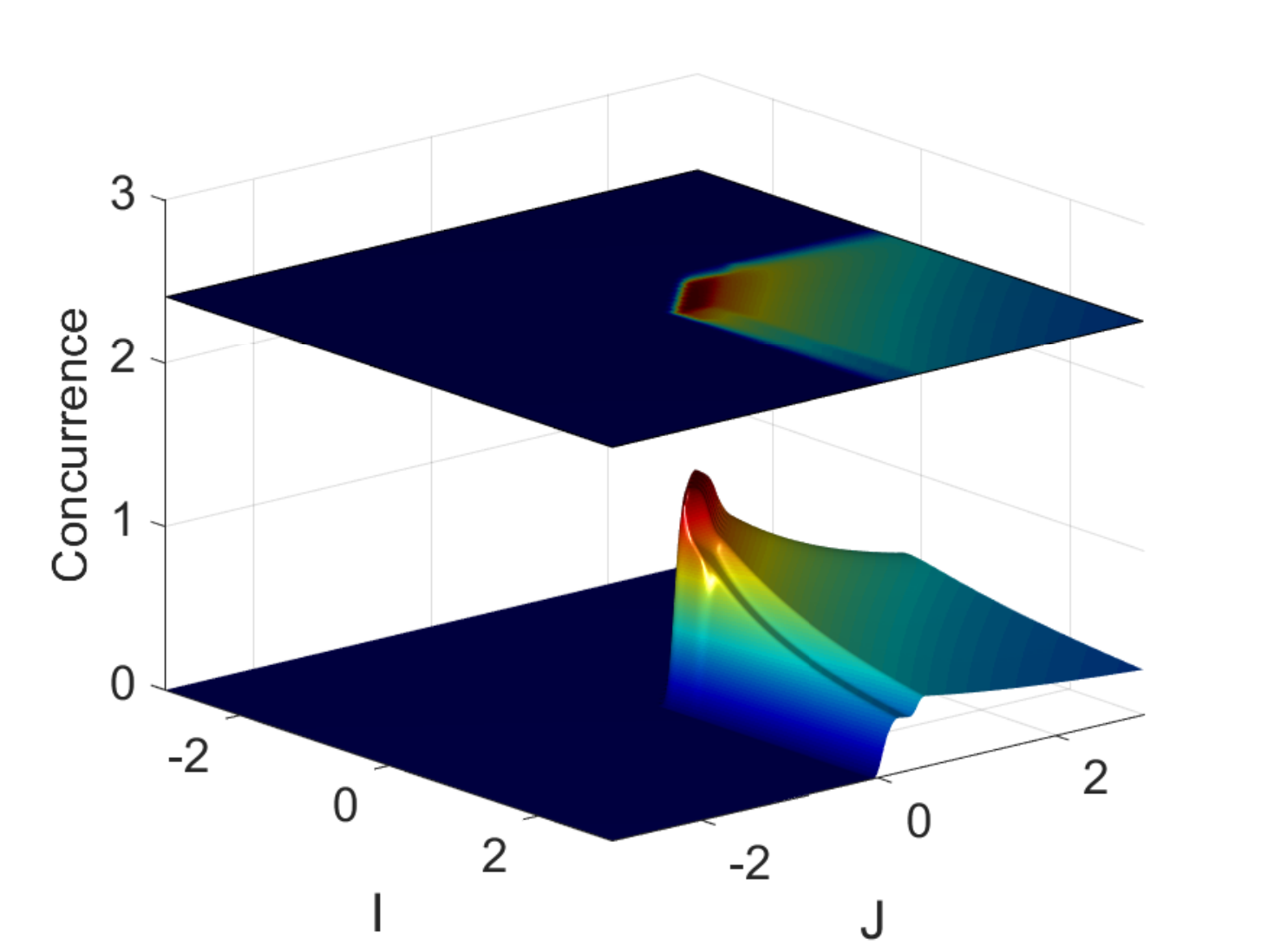}
(c) \includegraphics[width=0.43\columnwidth]{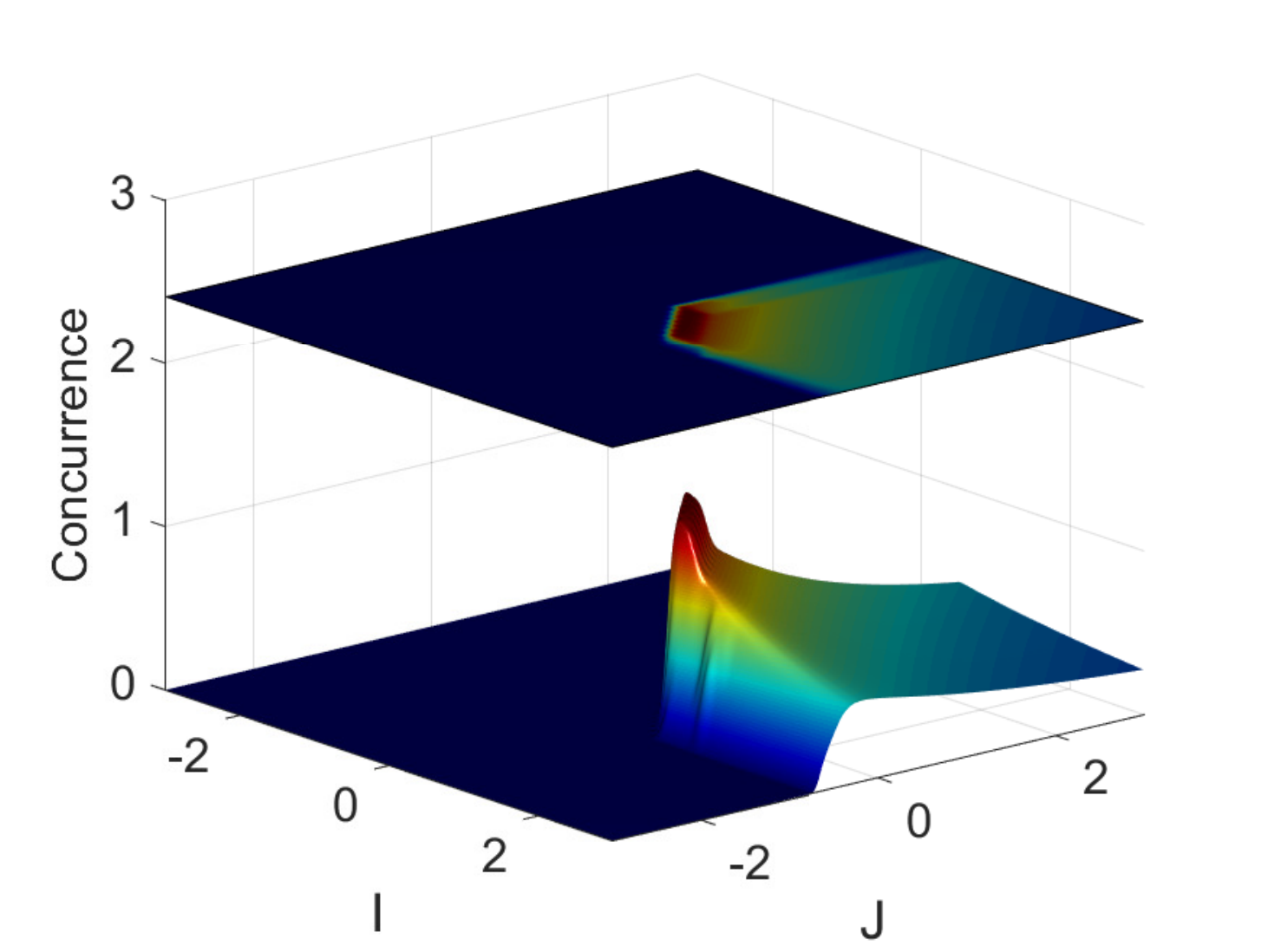}\\
(d) \includegraphics[width=0.43\columnwidth]{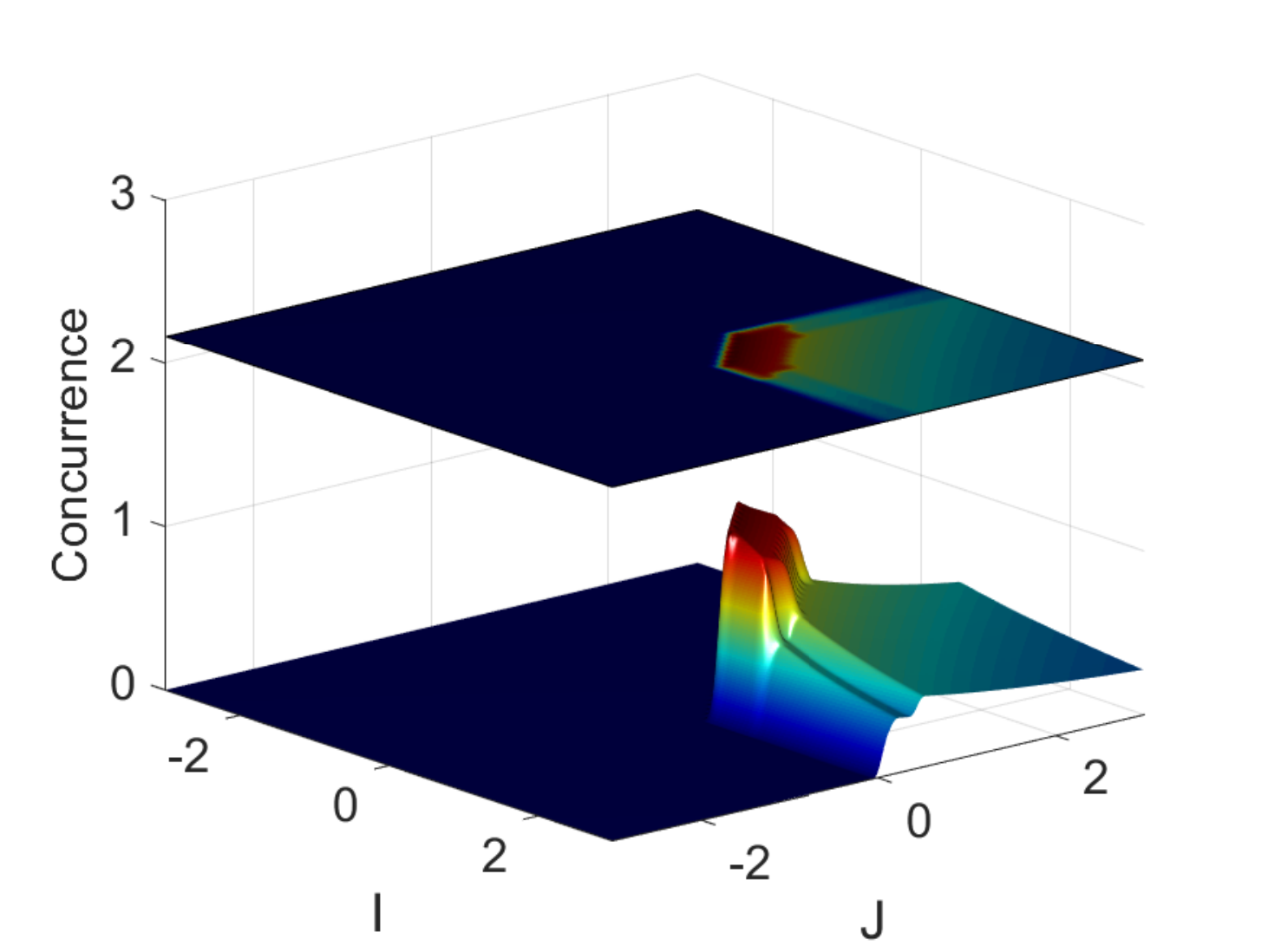}
(e) \includegraphics[width=0.43\columnwidth]{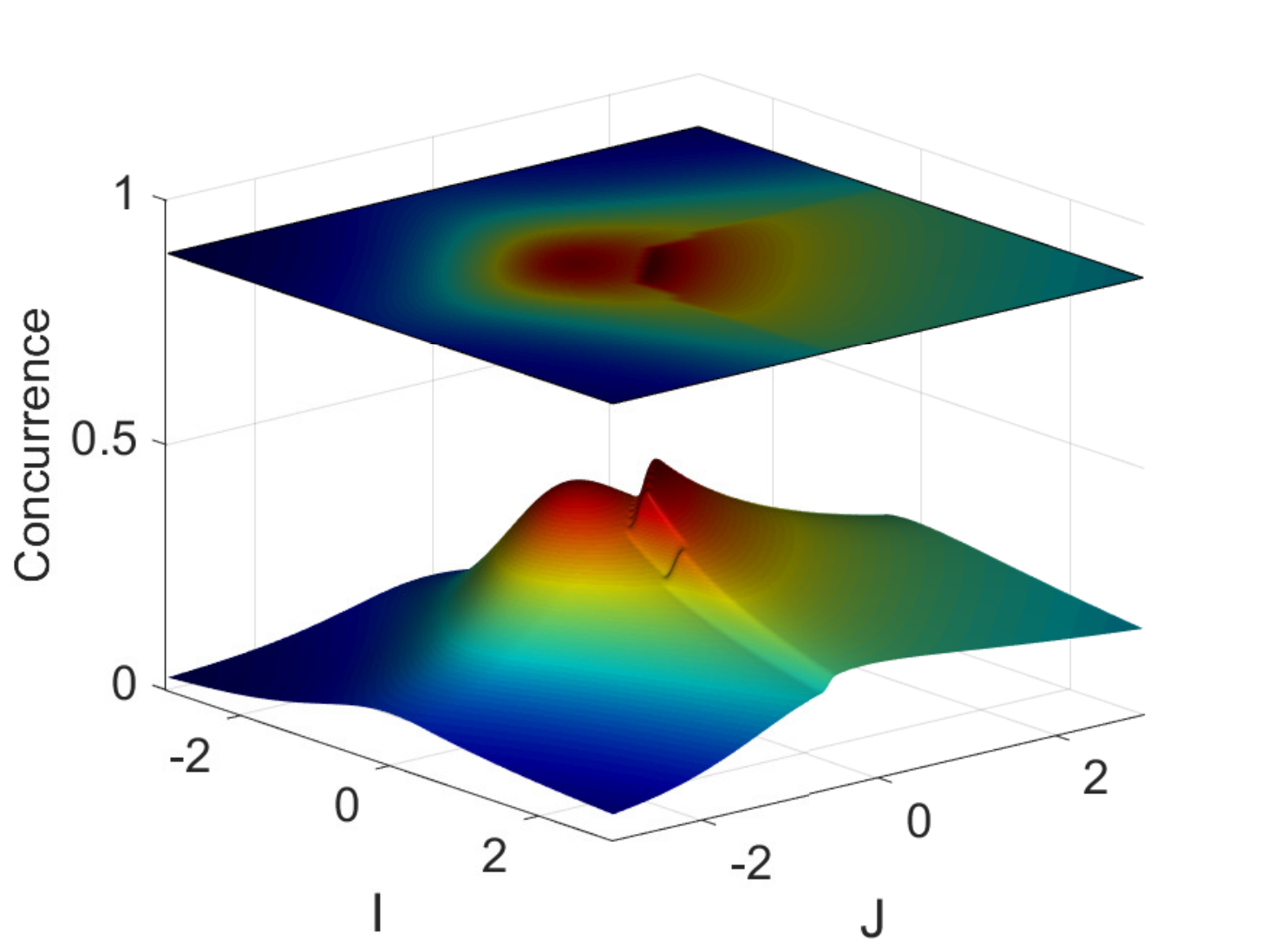}
\caption{Entanglement $C$\,(\ref{conc}) for intersubsystem exchange (\ref{eqKHTS1}) $\sim ( {\hat{\mathbf{S}}}_{\mathbf{i}} {\hat{\mathbf{S}}}_{\mathbf{j}}) ({\hat{\mathbf{T}}}_{\mathbf{i}} {\hat{\mathbf{T}}}_{\mathbf{j}})$
with negative $K=-1$ and external fields. In contrast to the case of $K = + 1$, here the $C$-maximum occurs not at a single point, but at a segment of the diagonal line.
(a) $\mathcal{H}_{s} = \mathcal{H}_{t} \ll 1$.
(b) $\mathcal{H}_{s} = 1, \mathcal{H}_{t} \ll 1$.
(c) $\mathcal{H}_{s} \ll 1, \mathcal{H}_{t} = 1$.
(d) $\mathcal{H}_{s} = 1$ and $\mathcal{H}_{t} = 1$.
(e) Staggered fields $|\mathcal{H}_{s}| = |\mathcal{H}_{t}| = 1$ in both subsystems.
Here, $\mathcal{H}_{s}$ and $\mathcal{H}_{t}$ stand for external fields in spin and pseudospin subsystems.
}
\label{K=-1_sstt}
\end{figure}

Fig.\,\ref{K=-1_sstt}a presents a measure of entanglement --- concurrence $C$\,(see Eq.\eqref{conc}) for negative intersubsystem exchange $K=-1$. As it can be expected, nonzero entanglement is observed in the area of a negative exchanges in both subsystems and its maximum is achieved for comparable values of $J$, $I$, and $K$. This acknowledges that not only the binding interaction $K$ between subsystems is decisive for the entanglement, but local interactions $J$ and $I$ as well. For $K > 0$ the same conclusion holds, see below.

\begin{figure}[t]
(a) \includegraphics[width=0.60\columnwidth]{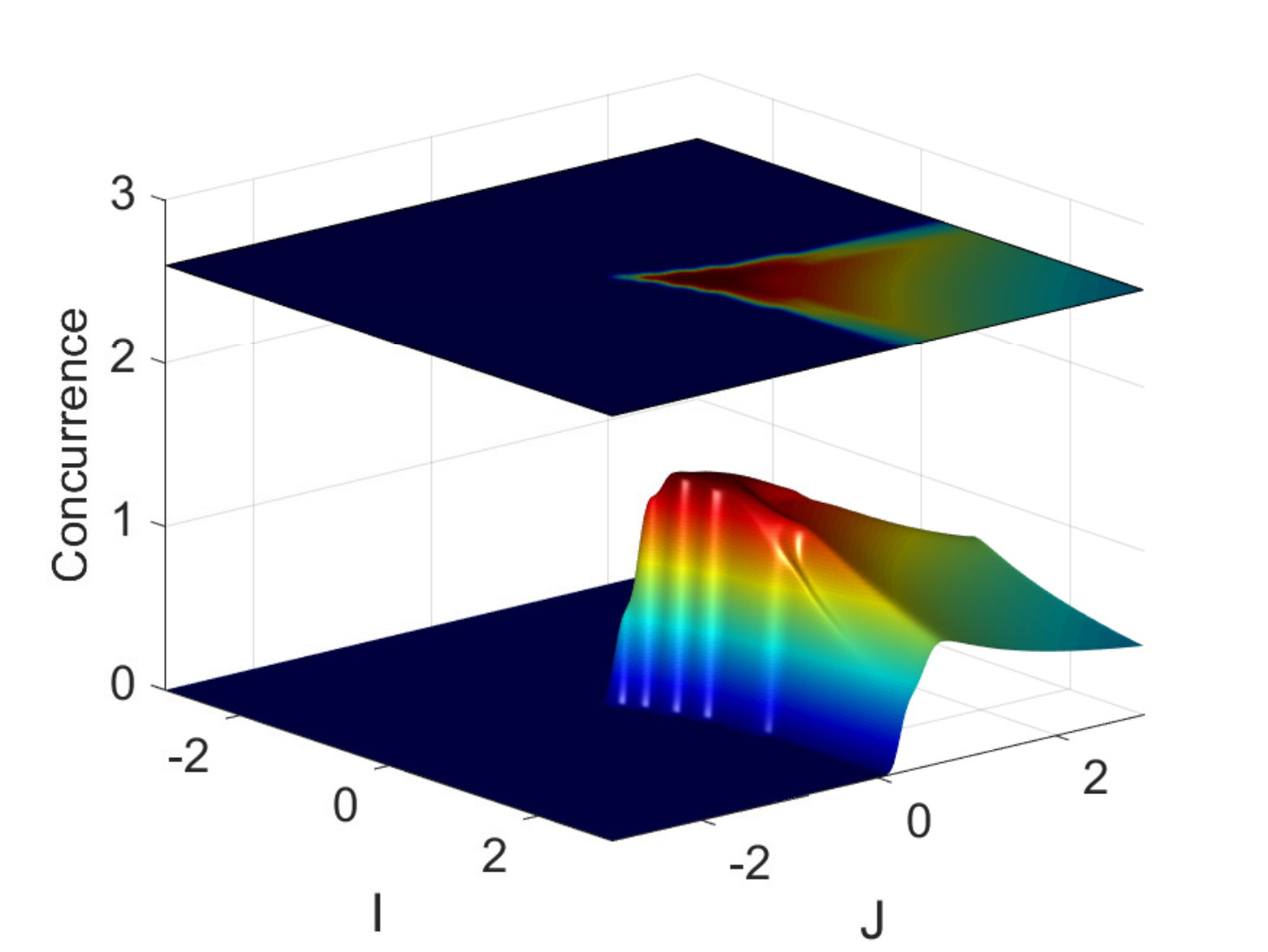}\\
(b) \includegraphics[width=0.43\columnwidth]{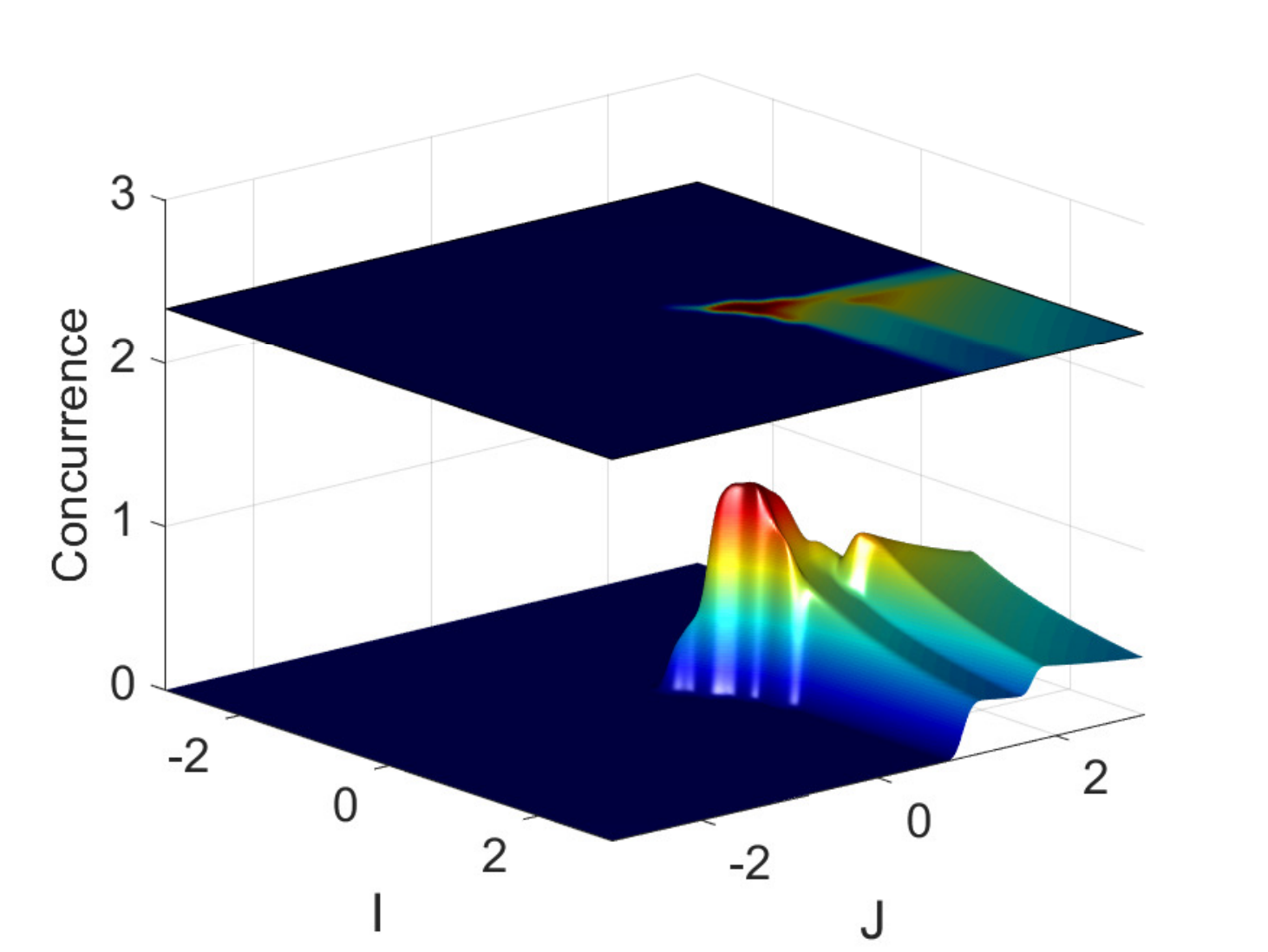}
(c) \includegraphics[width=0.43\columnwidth]{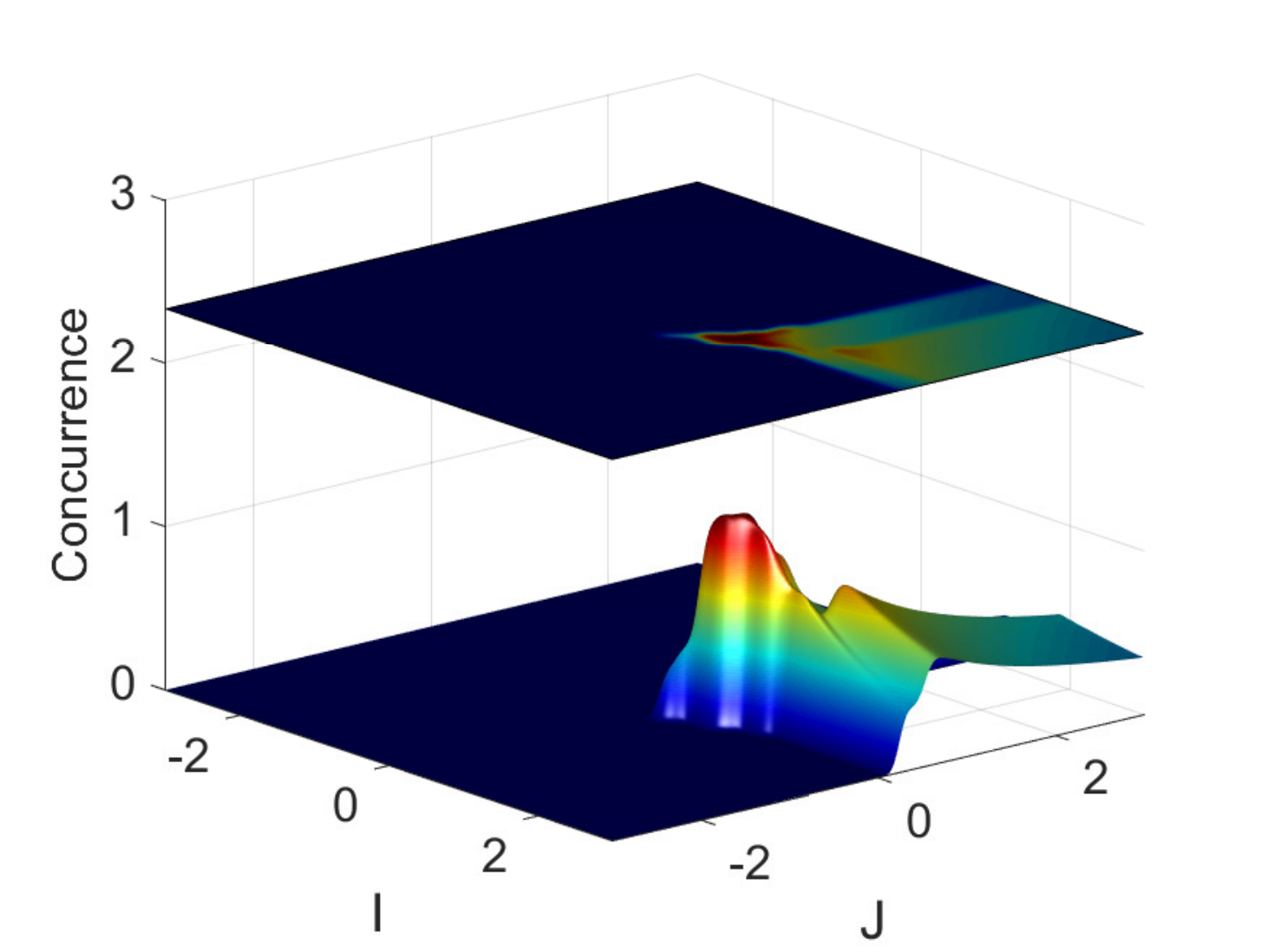}\\
(d) \includegraphics[width=0.43\columnwidth]{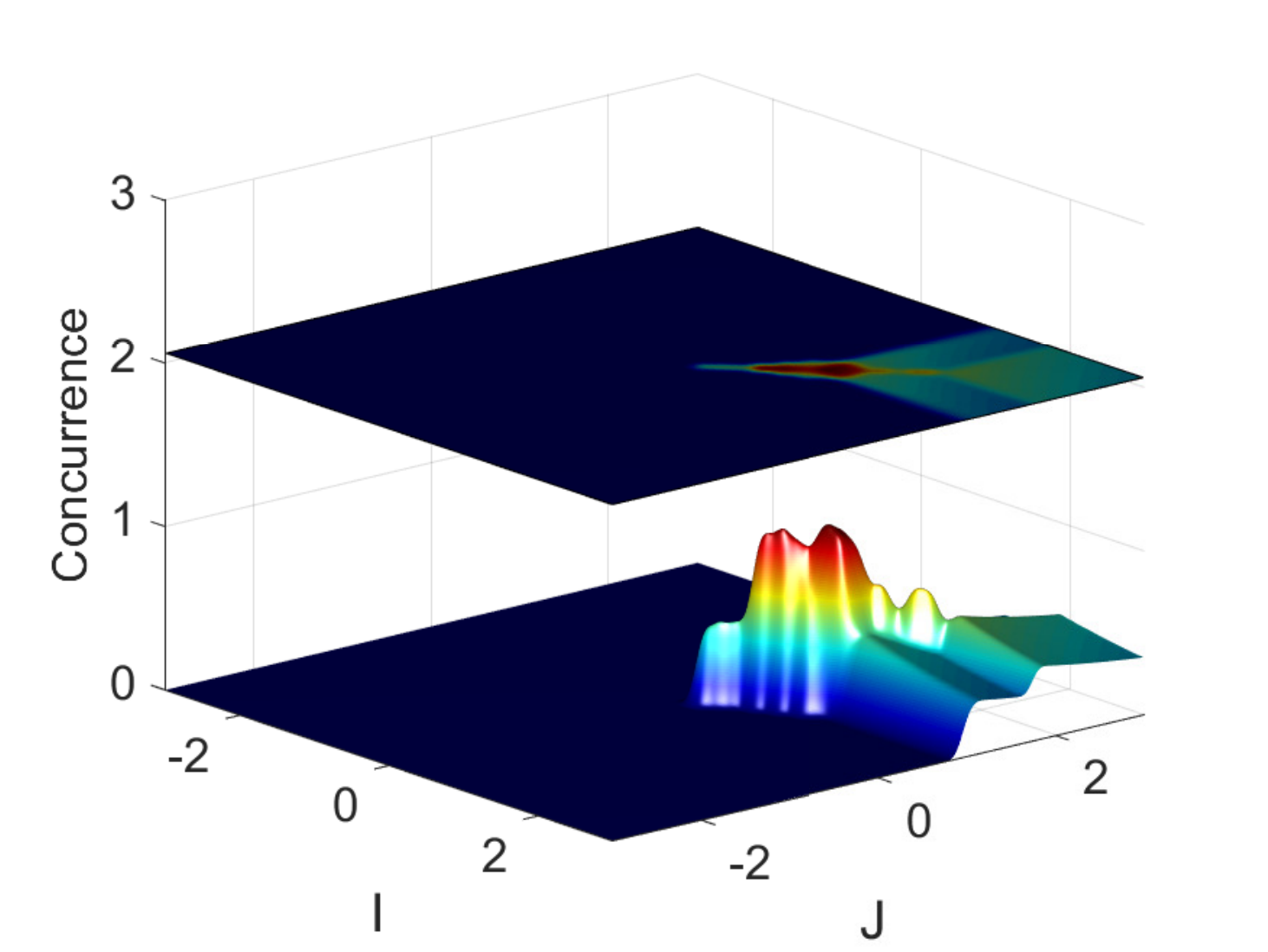}
(e) \includegraphics[width=0.43\columnwidth]{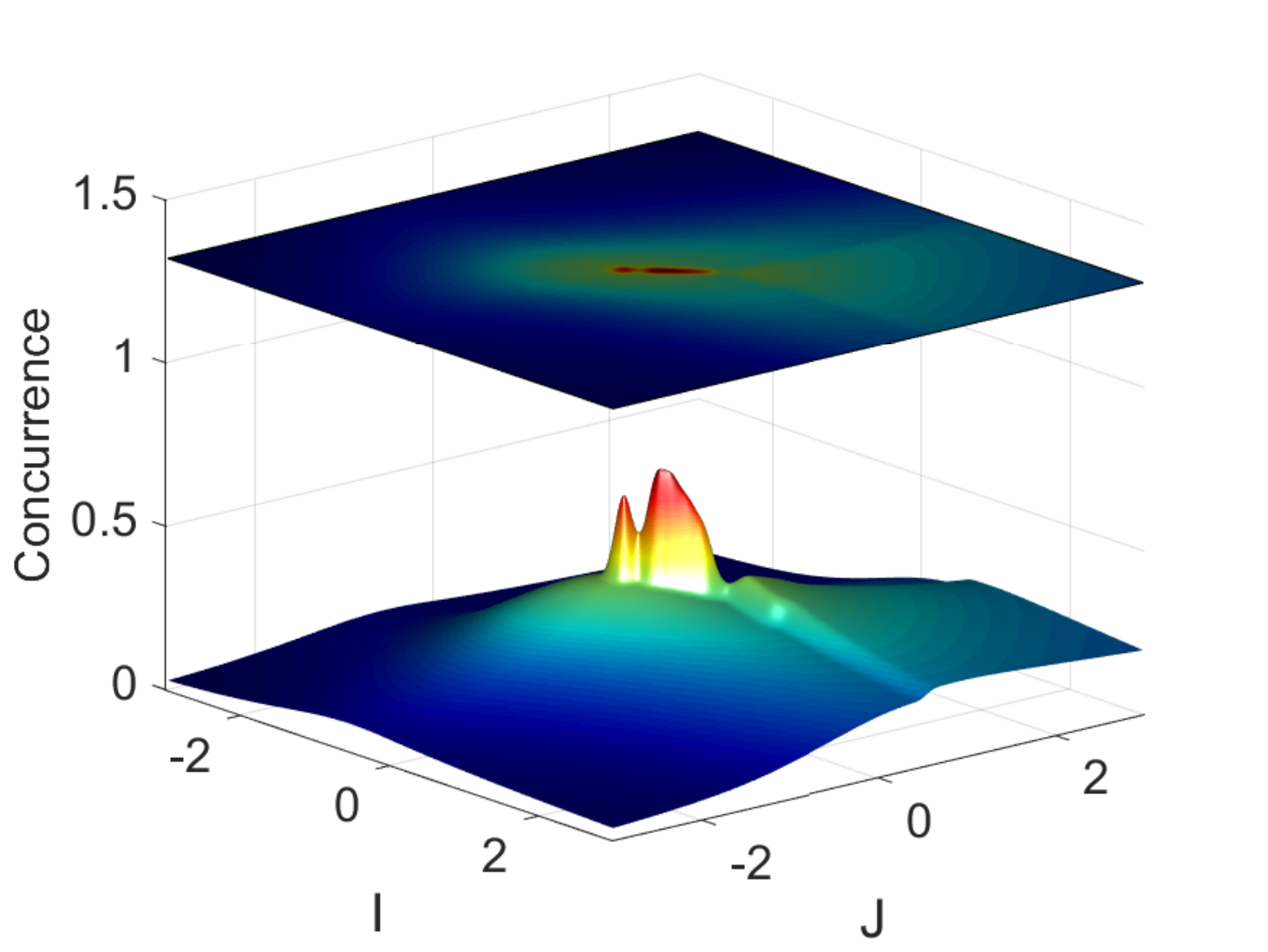}
\caption{Entanglement $C$\,(\ref{conc}) for intersubsystem exchange (\ref{eqKHTS1})
$\sim ( {\hat{\mathbf{S}}}_{\mathbf{i}} {\hat{\mathbf{S}}}_{\mathbf{j}}) ({\hat{\mathbf{T}}}_{\mathbf{i}} {\hat{\mathbf{T}}}_{\mathbf{j}})$
with positive $K = +1$. The maximum entanglement is achieved at single point.
(a) $\mathcal{H}_{s} = \mathcal{H}_{t} \ll 1$.
(b) $\mathcal{H}_{s} = 1, \mathcal{H}_{t} \ll 1$.
(c) $\mathcal{H}_{s} \ll 1, \mathcal{H}_{t} = 1$.
(d) $\mathcal{H}_{s} = 1$ and $\mathcal{H}_{t} = 1$.
(e) Staggered fields $|\mathcal{H}_{s}| = |\mathcal{H}_{t}| = 1$ in both subsystems.
Here, $\mathcal{H}_{s}$ and $\mathcal{H}_{t}$ stand for external fields in spin and pseudospin subsystems.
}
\label{K=+1_sstt}
\end{figure}

For $K=-1$, the phase (local order) boundary and the structure of the $C$-maximum differs significantly from the case $K=+1$ (see Fig.\,\ref{K=+1_sstt}a). Maximum of entanglement arises at a segment, while for $K=+1$ --- at a single point.

Figs.\,\ref{K=-1_sstt}a and \ref{K=+1_sstt}a qualitatively reproduce the known results~\cite{Chen07_PRB,You12_PRB}.
The spin-pseudospin structure in the finite entanglement area corresponds to AFM spin and AFM pseudospin local orders (this is supported by the intersubsystem local correlation functions, see Sec. S1 of Supplementary). Below in Sec.~\ref{sec:indicator}, we discuss the interconnection the entanglement and local correlators in depth, which is much less studied.

It appears that nonzero external fields change the degree of entanglement in different ways. We will discuss this in more detail in the foregoing Sec.~\ref{sec:fields}. Nevertheless, let us first note once more, that for both signs $K\gtrless 0$ the significant entanglement appears in the intuitive case of AFM exchanges in both subsystems, $J, I > 0$.

\subsection{\label{sec:fields} Entanglement and external fields}

We now discuss the effect of external fields on the entanglement. Let us note once again, that in the spin-orbital model different fields can be introduced in different subsystems, even if they act in the opposite directions.

It is intuitive that sufficiently large external field suppresses the entanglement, as it strengthens the tendency to form a common ferromagnetic state. We will discuss below that the entanglement area transformation under strong external field is not so trivial, especially in a very frustrated case $J \sim I \sim K$, where entanglement
typically has a maximum.

With the negative sign of the intersubsystem exchange $K=-1$, the initial zero-field picture under the influence of an external field shifts, almost without deformation, along the corresponding coordinate axis, see  Figs.\,\ref{K=-1_sstt}b--c.

The case of two simultaneously acting fields is more peculiar. A local area of strong entanglement is formed, having a tooth-shape, see Fig.\,\ref{K=-1_sstt}d. The result is practically independent of the mutual orientation of the fields.
With magnification of the field amplitude, the localization effect increases, though is not transformed qualitatively, so we put the corresponding figure to Sec. S2 of Supplementary.

The effect of the staggered fields (similar in both subsystems) is even more amazing, see Fig.\,\ref{K=-1_sstt}e. The area substantial entanglement in $J$--$I$ plane is dramatically enlarged and the non-zero entanglement appears in the domains, where it was negligible in all other cases under discussion.

\begin{figure*}[tbp]
  \centering
  \mbox{
    \subfigure[Spin-pseudospin single-site correlator] {\includegraphics[width=0.3\textwidth]{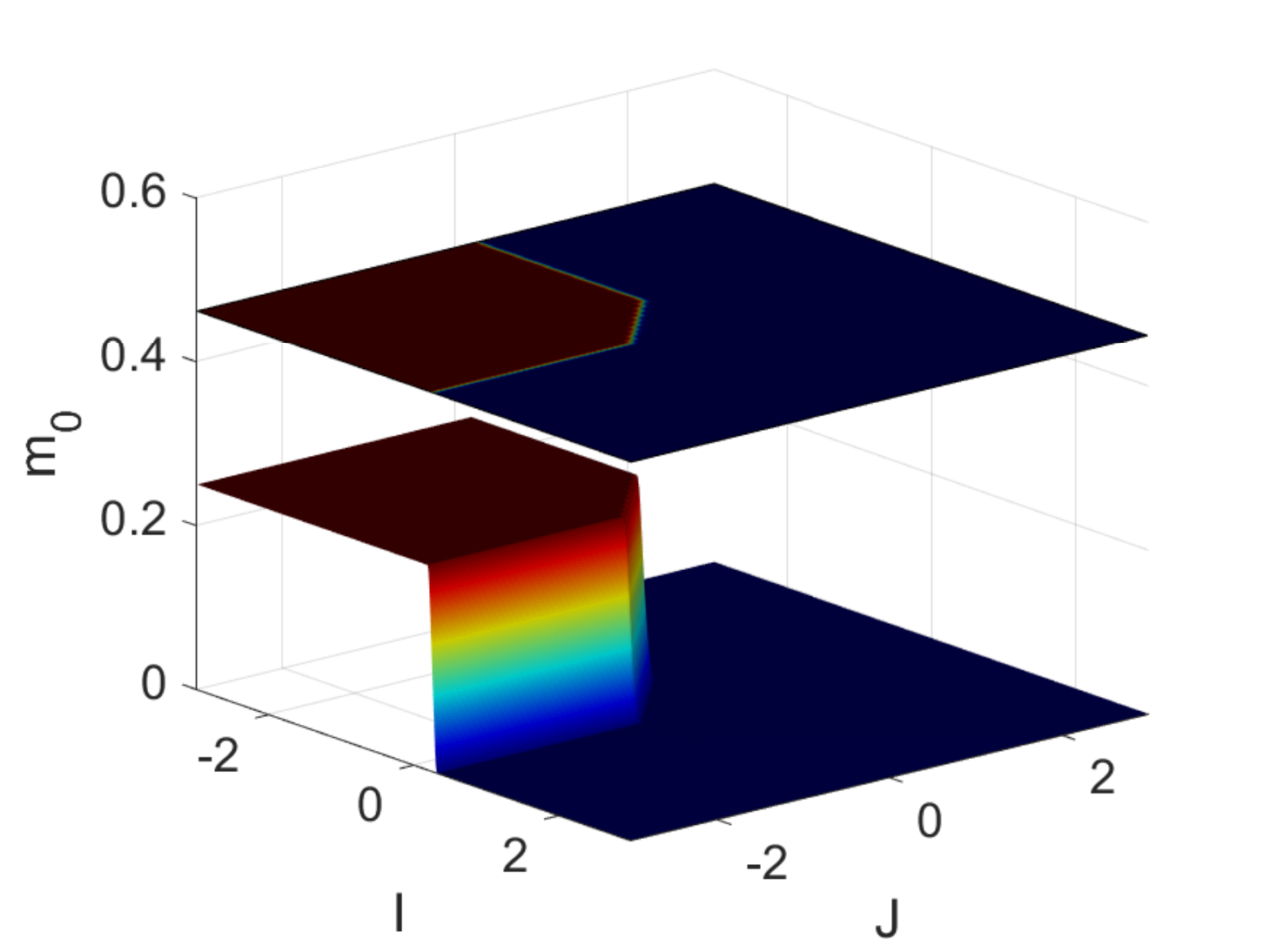}}\quad
    \subfigure[Spin-pseudospin two-site correlator] {\includegraphics[width=0.3\textwidth]{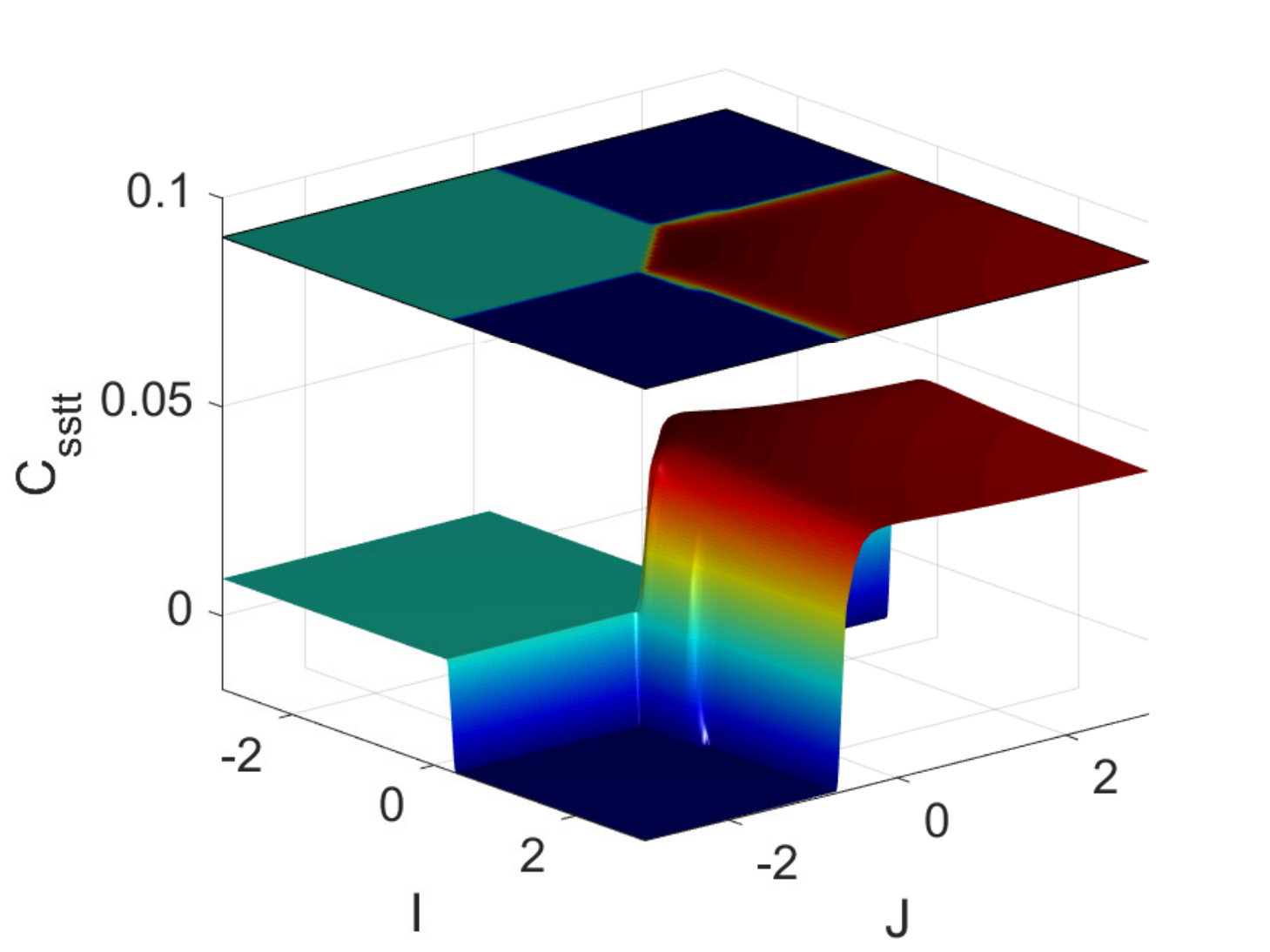}}\quad
    \subfigure[Gradient of the two-site correlator] {\includegraphics[width=0.27\textwidth]{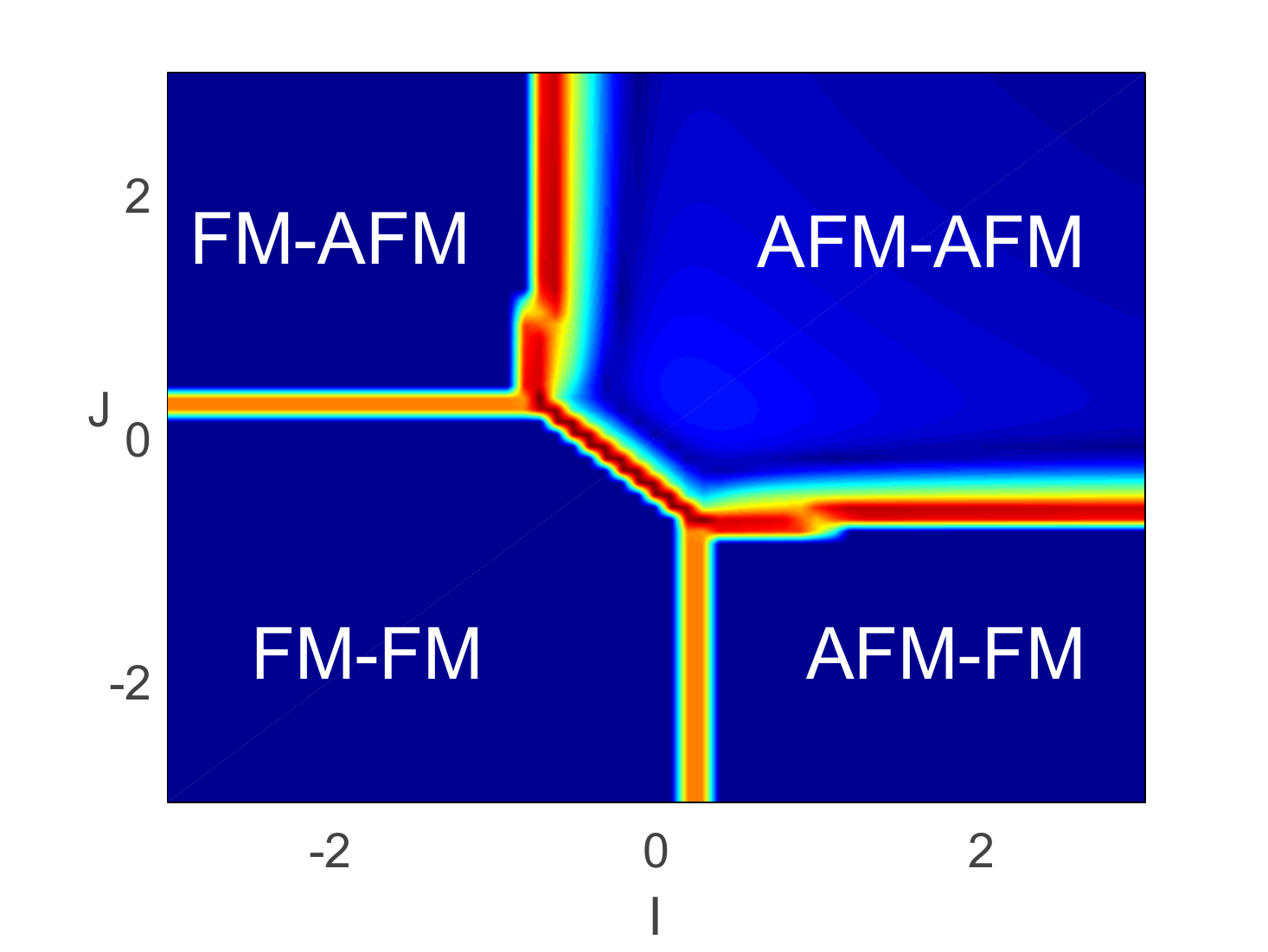}}
  }
  \caption{ $K =-1$
  (a) The chain-averaged single-site spin-pseudospin correlator does not provide accurate information on the entanglement.
  (b) The two-site spin-pseudospin correlator reveals the entanglement region boundaries.
  (c) The gradient of the two-site correlator allows us to select the area of entanglement. The order structure in spin and pseudospin subsystem is designated.
  }
  \label{K=-1_Indicate}
\end{figure*}

\begin{figure*}
  \centering
  \mbox{
    \subfigure[Spin-pseudospin single-site correlator] {\includegraphics[width=0.3\textwidth]{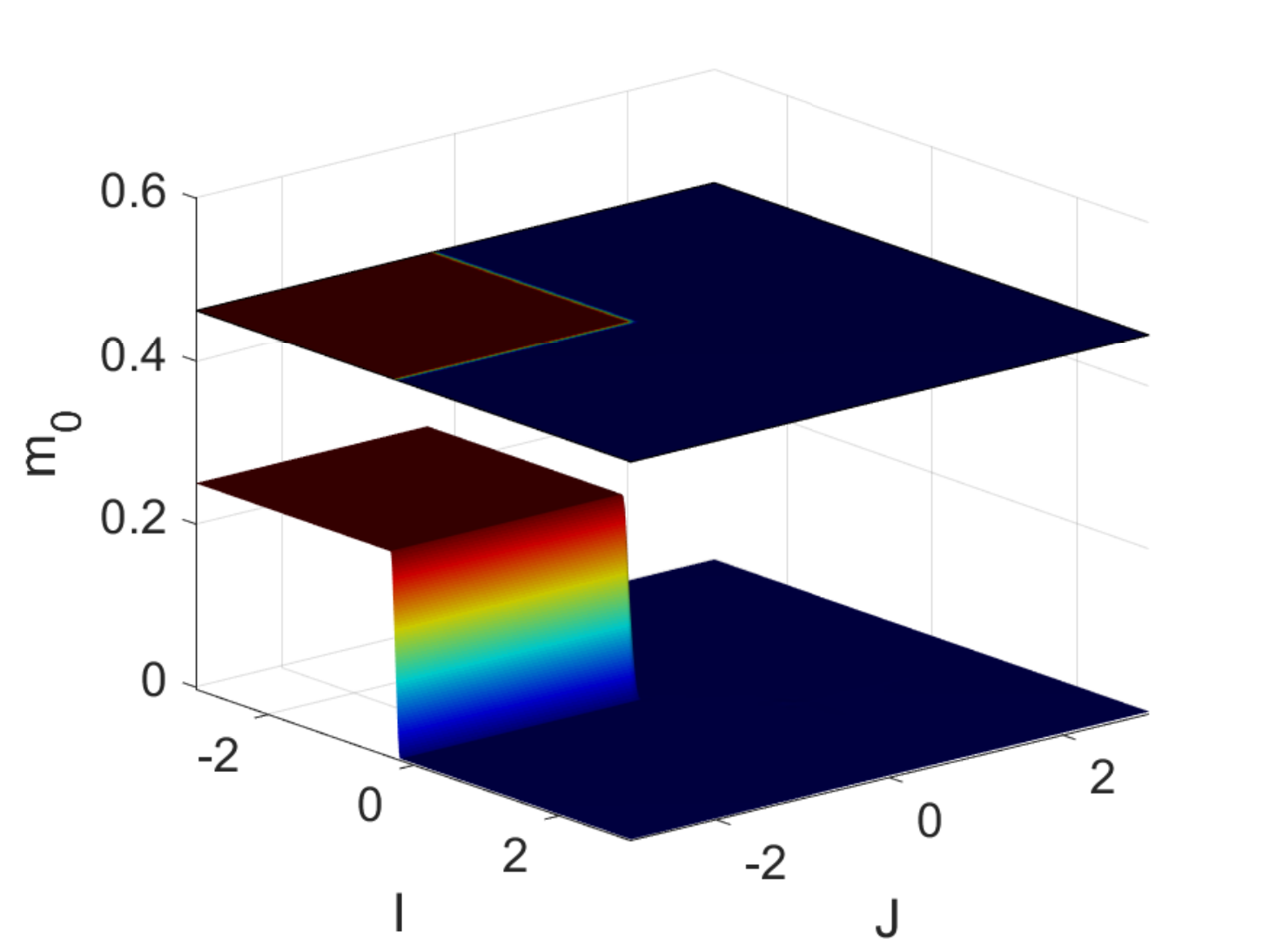}}\quad
    \subfigure[Spin-pseudospin two-site correlator] {\includegraphics[width=0.3\textwidth]{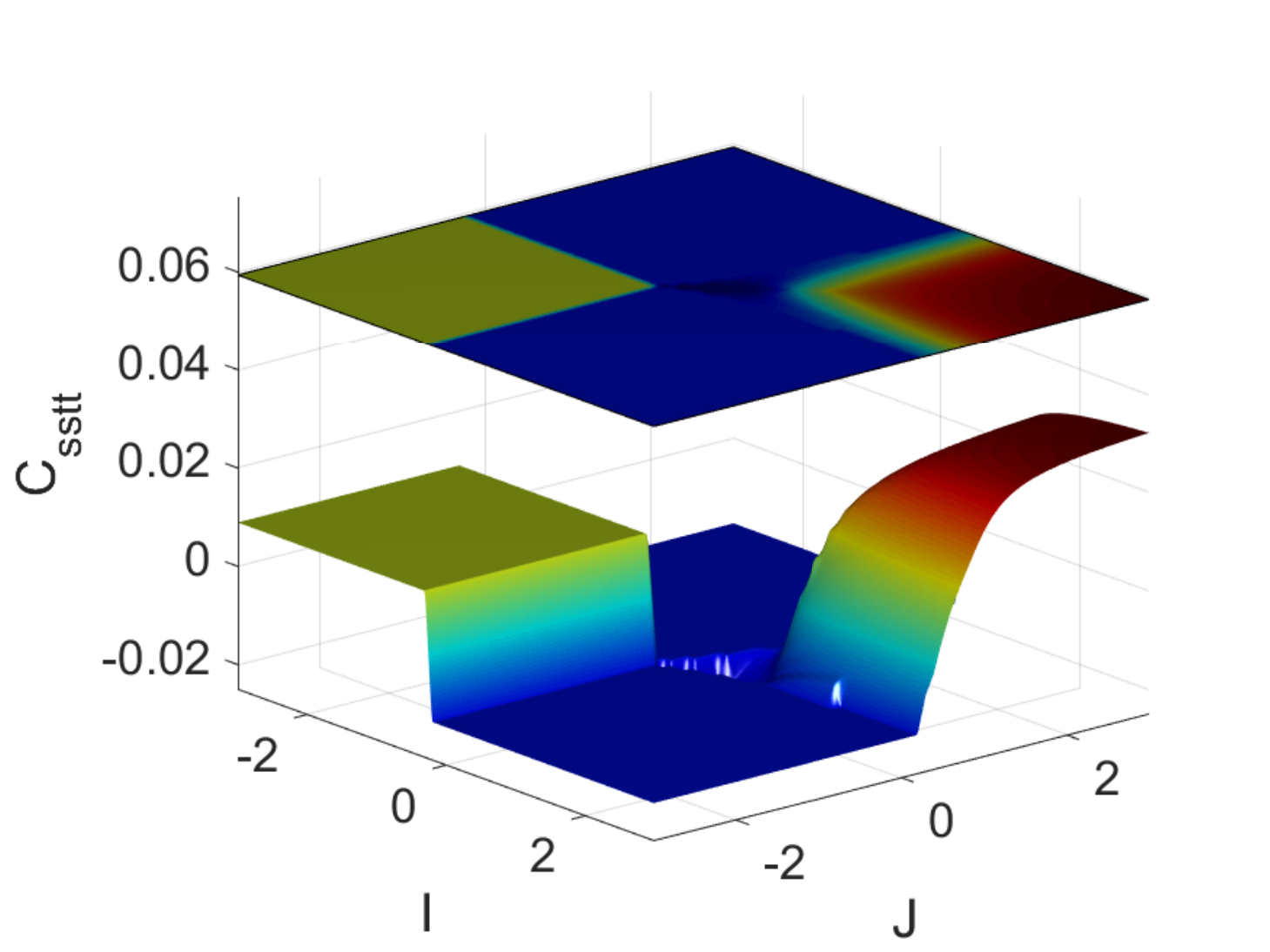}}\quad
    \subfigure[Gradient of the two-site correlator] {\includegraphics[width=0.27\textwidth]{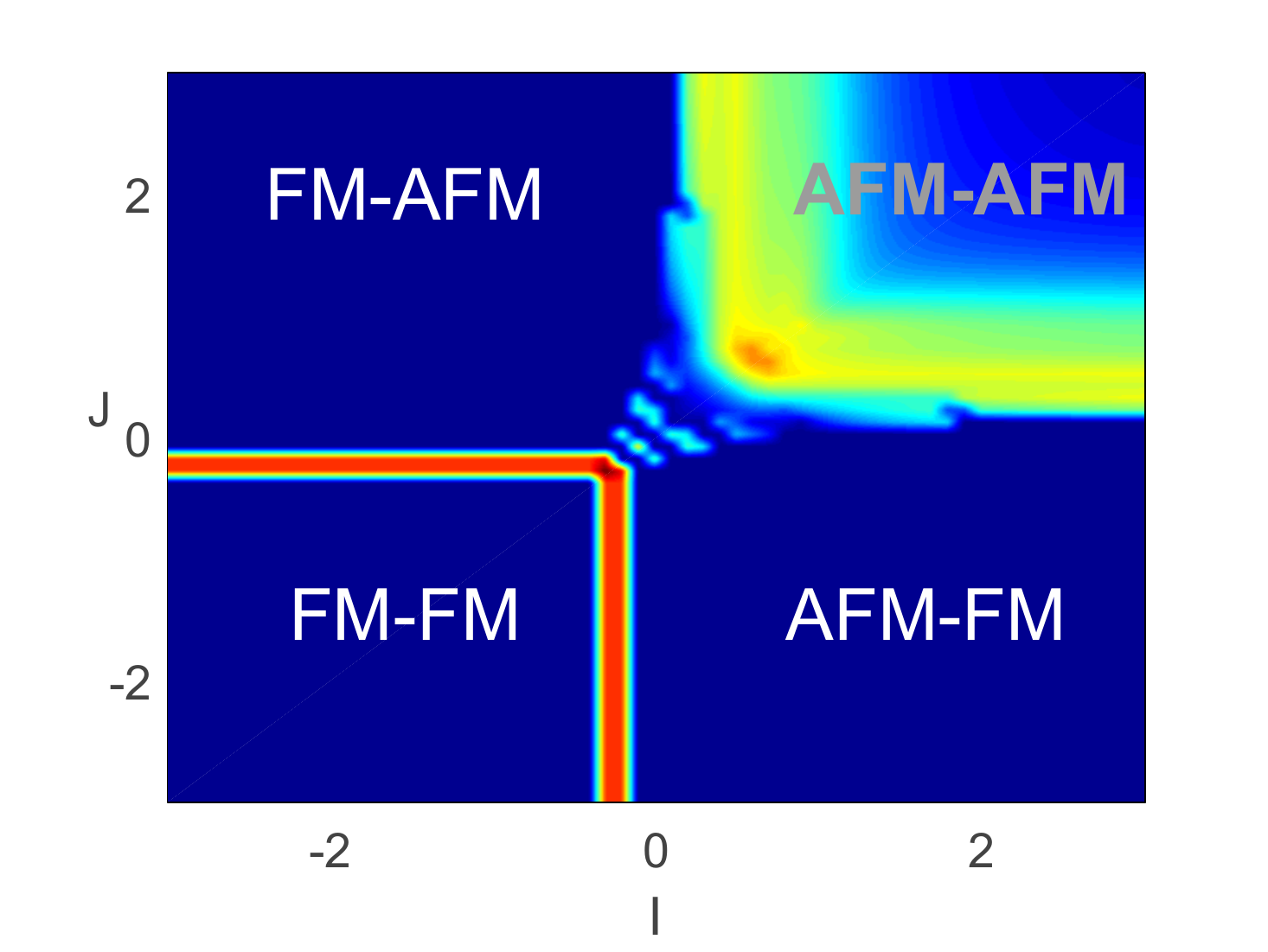}}
  }
  \caption{$K = +1$. Analogous to Fig.\,\ref{K=-1_Indicate}
  (a) The chain-averaged single-site spin-pseudospin correlator does not provide accurate information on the entanglement.
  (b) The two-site spin-pseudospin correlator detects reveals the entanglement region boundaries.
  (c) The gradient of the two-site correlator allows us to select the area of entanglement. The order structure in spin and pseudospin subsystem is designated.
  }
  \label{K=+1_Indicate}
\end{figure*}

Similar effects are observed with the positive sign of intersubsystem exchange $K=+1$. Here, also, the initial nullipole pattern under an external field shifts along the corresponding coordinate axis; see the Fig.\,\ref{K=+1_sstt}b-c. Nevertheless, some deformation of the initial structure are observed with a nonmonotonic behavior of entanglement with increasing $J$ or $I$.

With two simultaneously nonzero fields, as well as for $K=-1$, the local area of entanglement is formed, see Fig.\,\ref{K=+1_sstt}d. Moreover, the situation is almost unrelated to the mutual orientation of the fields, and as the field amplitude increases, the localization effect of a region of strong entanglement becomes more pronounced.

The destructive effect of the staggered fields in the case of positive
$K = +1$ is much stronger, than for $K = -1$, see Fig.\,\ref{K=+1_sstt}e.
Only sharp narrow segment near
$|J| \sim |I| \lesssim 1$ survives against the smooth concurrence background.

\section{\label{sec:indicator} Interrelation between spin-pseudospin correlation functions and entanglement}

Correlation functions between spin (pseudospin) degrees of freedom provide important information about the system state structure. On one hand, you can find out the local structure in the spin and pseudospin subspace that allows us to roughly distinguish ``FM'' and ``AFM''-like local ordering. On the other hand, irreducible intersubsystem correlators may be sensitive to entanglement effects. We are studying this question in detail below.

It would be natural to expect a one-site intersubsystem correlator $\langle {\hat{\mathbf{S}}}_{\mathbf{i}} {\hat{\mathbf{T}}}_{\mathbf{i}}\rangle$ to be related with entanglement. Nevertheless, our analysis shows that in the general case, a chain-averaged single-site spin-pseudospin correlator does not provide accurate information on the entanglement region. As an example see Fig.\,\ref{K=-1_Indicate}a in comparison with Figs.\,\ref{K=-1_sstt}a: the single-site spin-pseudospin correlator reproduces only one small segment of the entanglement area borders (this is also the case for other types of intersystem interaction considered posterior).

Thus, one should address a two-site correlator:
$\langle {\hat{\mathbf{S}}}_{\mathbf{i}}{\hat{\mathbf{S}}}_{\mathbf{j}} {\hat{\mathbf{T}}}_{\mathbf{i}}{\hat{\mathbf{T}}}_{\mathbf{j}}\rangle$.
In Fig.\,\ref{K=-1_Indicate}b, two-site spin-pseudospin correlator
$\langle {\hat{\mathbf{S}}}_{\mathbf{i}}{\hat{\mathbf{S}}}_{\mathbf{j}} {\hat{\mathbf{T}}}_{\mathbf{i}}{\hat{\mathbf{T}}}_{\mathbf{j}}\rangle$
($\mathbf{i},\mathbf{j}$ are the nearest neighbors) is shown for $K=-1$. According to Fig.\,\ref{K=-1_sstt}a, the two-site spin-pseudospin correlator reasonably reproduces the entanglement region boundaries.
Note, that one-site and two-site irreducible correlators (covariances)
$\langle {\hat{\mathbf{S}}}_{\mathbf{i}} {\hat{\mathbf{T}}}_{\mathbf{i}}\rangle -
\langle {\hat{\mathbf{S}}}_{\mathbf{i}}\rangle
\langle{\hat{\mathbf{T}}}_{\mathbf{i}}\rangle$,
$\langle {\hat{\mathbf{S}}}_{\mathbf{i}}{\hat{\mathbf{S}}}_{\mathbf{j}}
{\hat{\mathbf{T}}}_{\mathbf{i}}{\hat{\mathbf{T}}}_{\mathbf{j}}\rangle -
\langle {\hat{\mathbf{S}}}_{\mathbf{i}}{\hat{\mathbf{S}}}_{\mathbf{j}}\rangle \langle
{\hat{\mathbf{T}}}_{\mathbf{i}}{\hat{\mathbf{T}}}_{\mathbf{j}}\rangle $
lead to the same result for phase boundaries as the initial correlators.

A much clearer picture of the boundaries is visible in Fig.\,\ref{K=-1_Indicate}c, where the gradient (in the parameters' space) of a two-site spin-pseudospin correlator is presented. The gradient structure allows one to distinguish the entanglement in the phase diagram precisely.

In Fig.\,\ref{K=+1_Indicate}, similar data are shown for positive spin-pseudospin exchange $K = + 1$.
In Fig.\,\ref{K=+1_Indicate}a, the single-site spin-pseudo-spin correlator does not provide information on the entanglement area
(compare with Fig.\,\ref{K=+1_sstt}a).
In Fig.\,\ref{K=+1_Indicate}b similarly to Fig.\,\ref{K=-1_Indicate}b two-site spin-pseudo-spin correlator reasonably reproduces boundaries of the entanglement region.
Finally, the gradient of a two-site spin-pseudospin correlator
selects the entanglement --- Fig.\,\ref{K=+1_Indicate}c accurately.

\begin{figure}[b]
(a) \includegraphics[width=0.60\columnwidth]{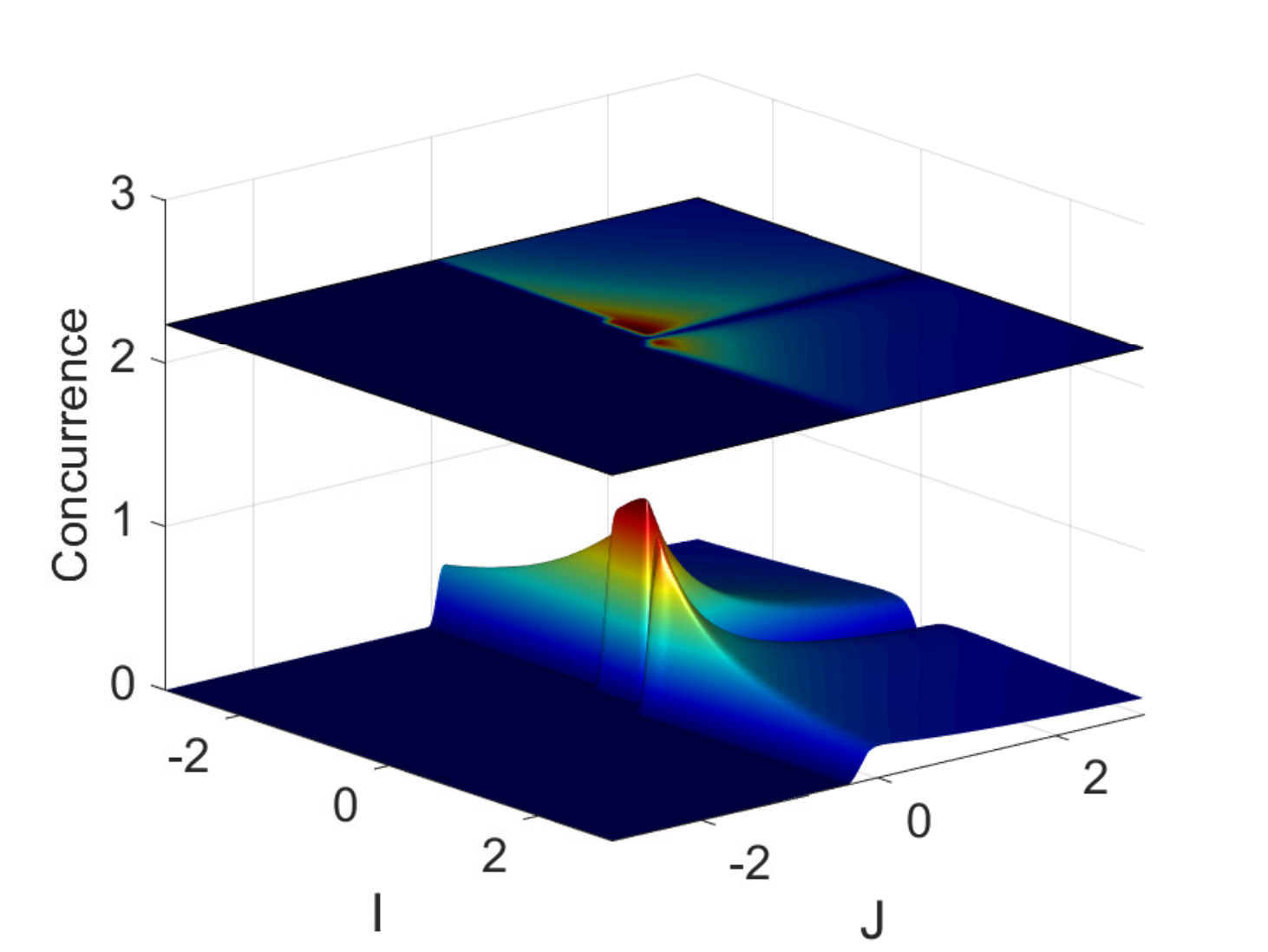} \\
(b) \includegraphics[width=0.43\columnwidth]{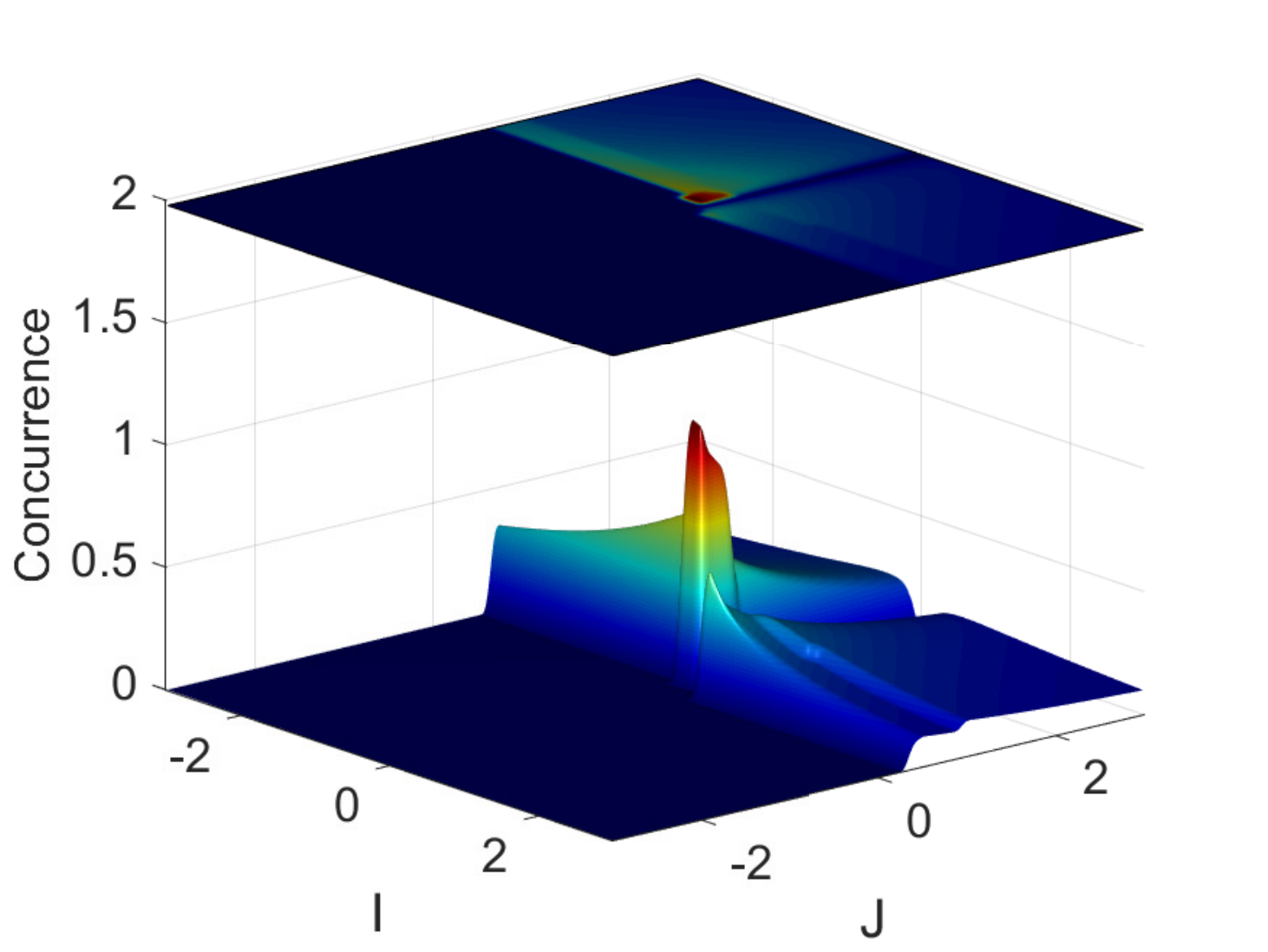}
(c) \includegraphics[width=0.43\columnwidth]{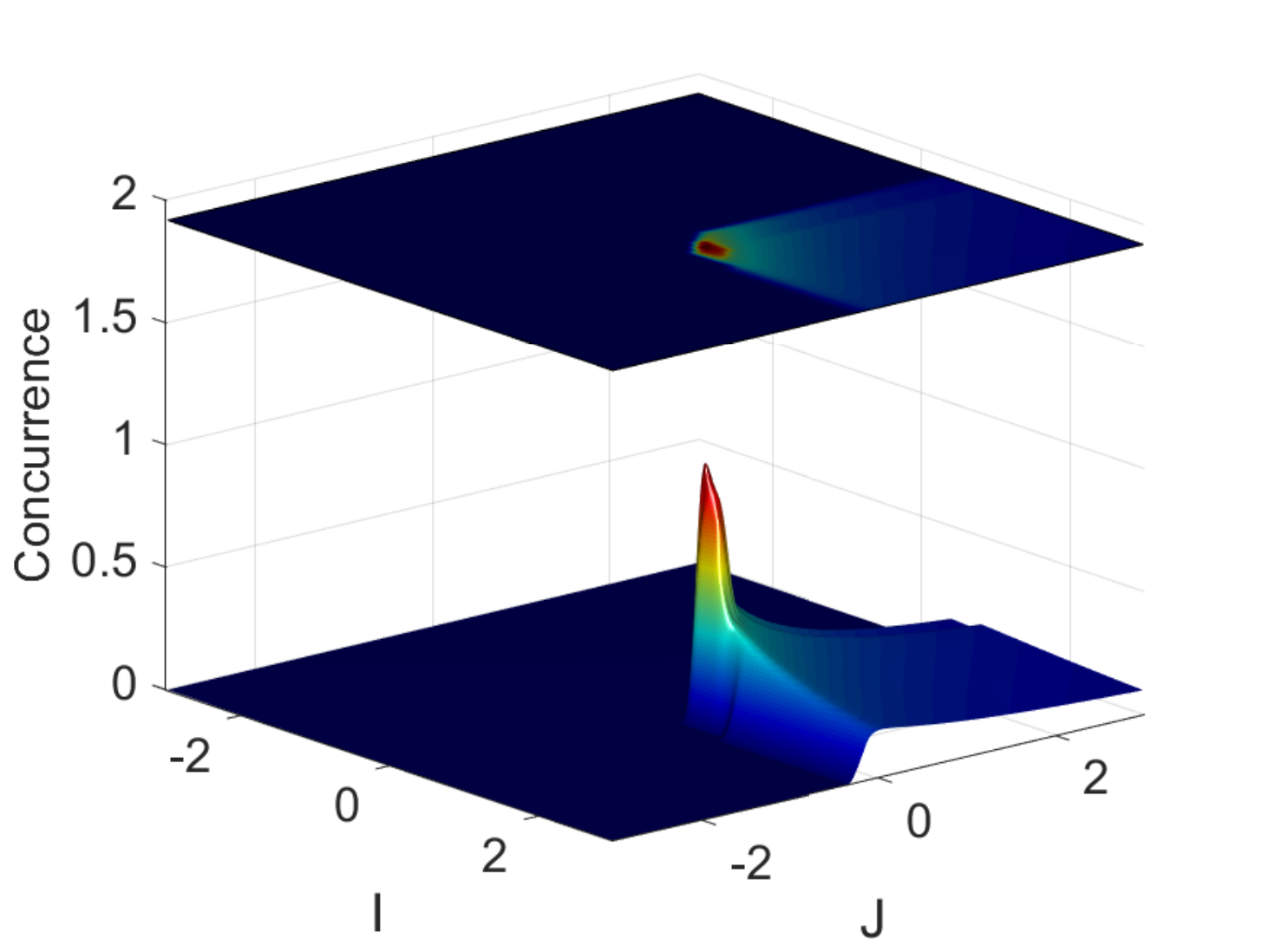}\\
(d) \includegraphics[width=0.43\columnwidth]{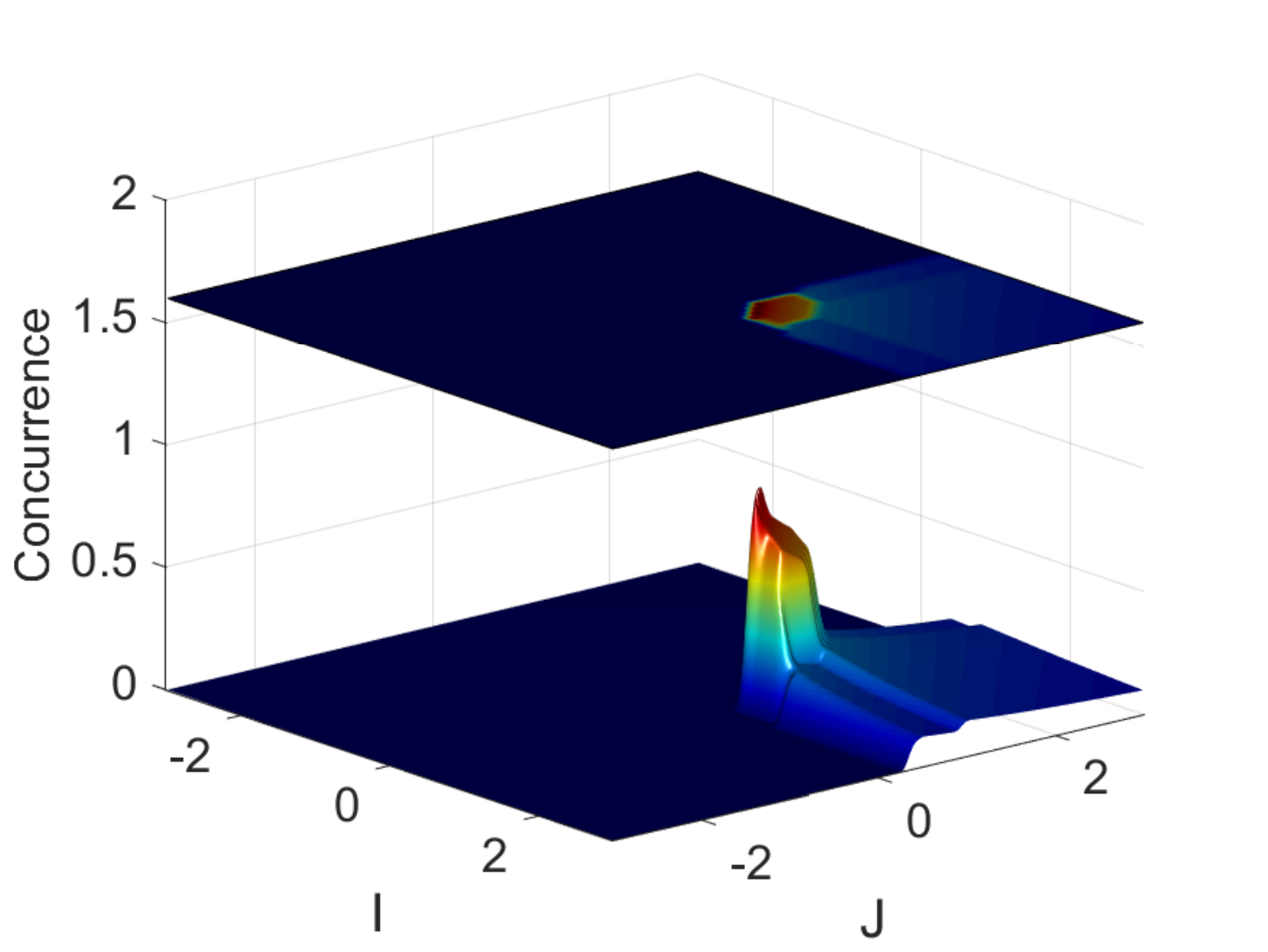}
(e) \includegraphics[width=0.43\columnwidth]{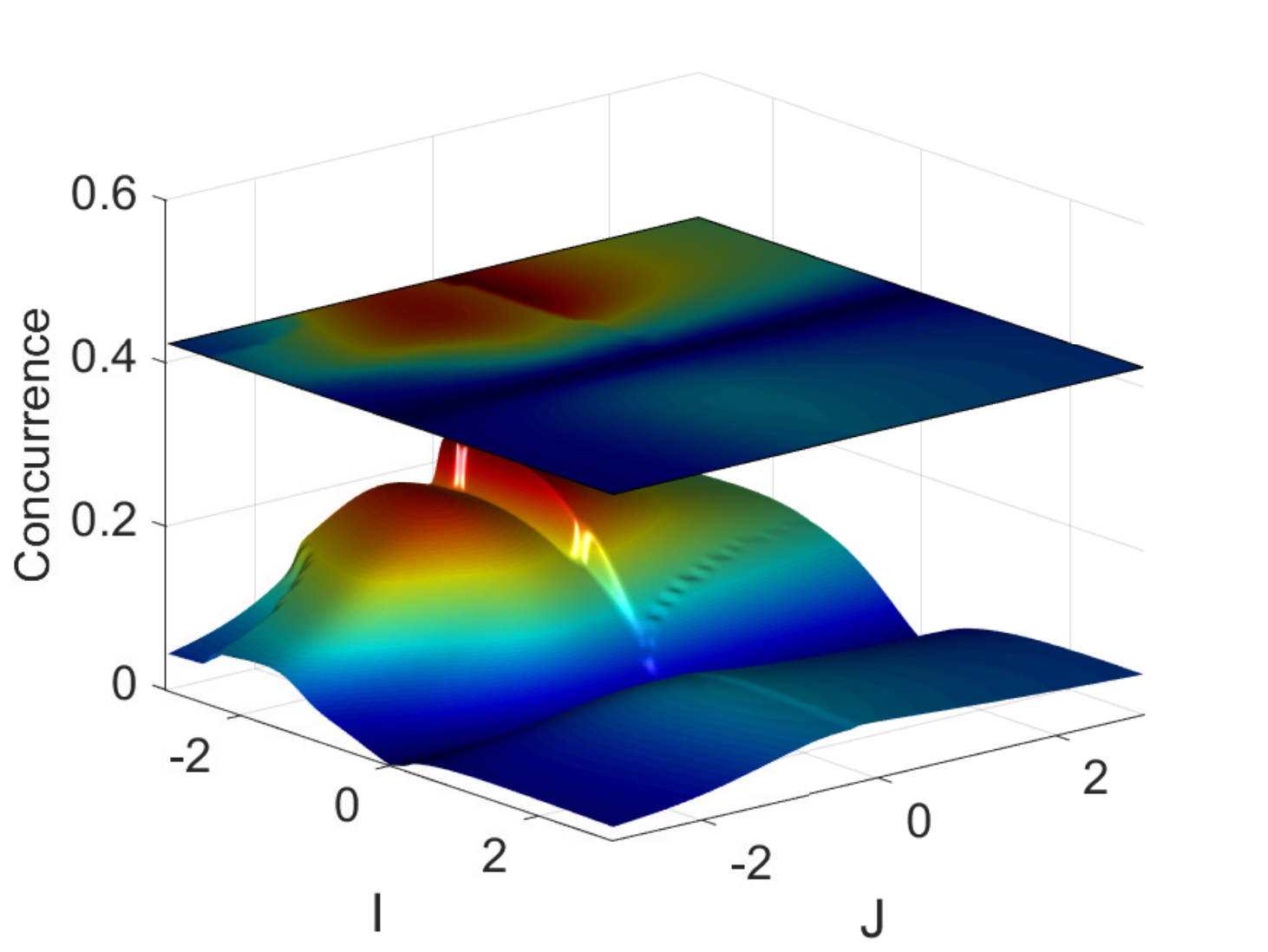}
\caption{Entanglement $C$\,(\ref{conc}) for intersubsystem exchange (\ref{eqKHTS2})
$ \sim \left( {\hat{\mathbf{S}}}_{\mathbf{i}} {\hat{\mathbf{S}}}_{\mathbf{j}}\right) \left(T_{\mathbf{i}}^{z}T_{\mathbf{j}}^{z}\right)$
with negative $K=-1$. At zero external fields, the entangled state is realized at a half of the phase plane.
(a) $\mathcal{H}_{s} = \mathcal{H}_{t} \ll 1$.
(b) $\mathcal{H}_{s} = 1, \mathcal{H}_{t} \ll 1$.
(c) $\mathcal{H}_{s} \ll 1, \mathcal{H}_{t} = 1$.
(d) $\mathcal{H}_{s} = 1$ and $\mathcal{H}_{t} = 1$.
(e) Staggered fields $|\mathcal{H}_{s}| = |\mathcal{H}_{t}| = 1$ in both subsystems.
Here, $\mathcal{H}_{s}$ and $\mathcal{H}_{t}$ stand for external fields in spin and pseudospin subsystems.
}
\label{K=-1_sszz}
\end{figure}
\begin{figure}[b]
(a) \includegraphics[width=0.60\columnwidth]{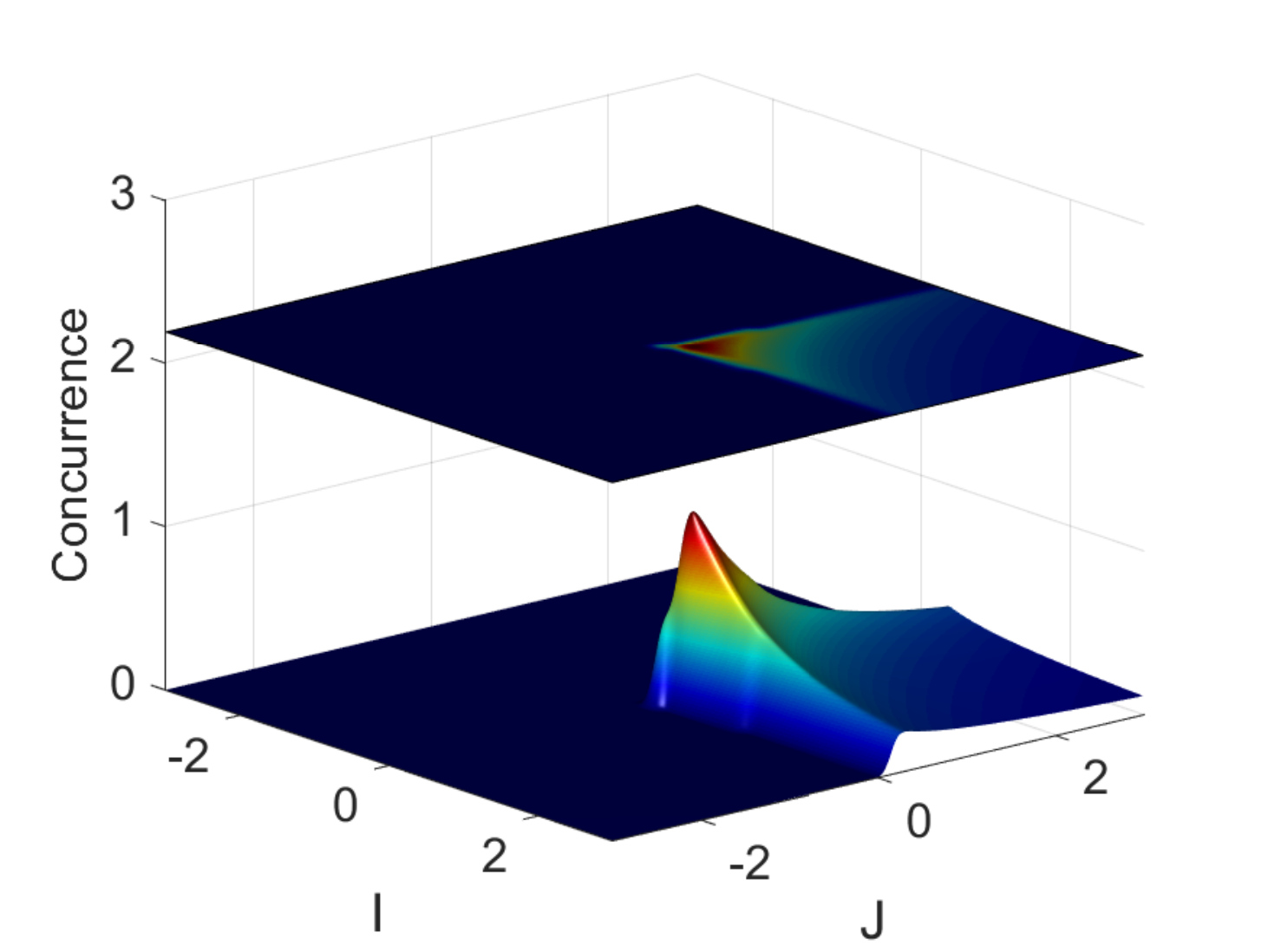}\\
(b) \includegraphics[width=0.43\columnwidth]{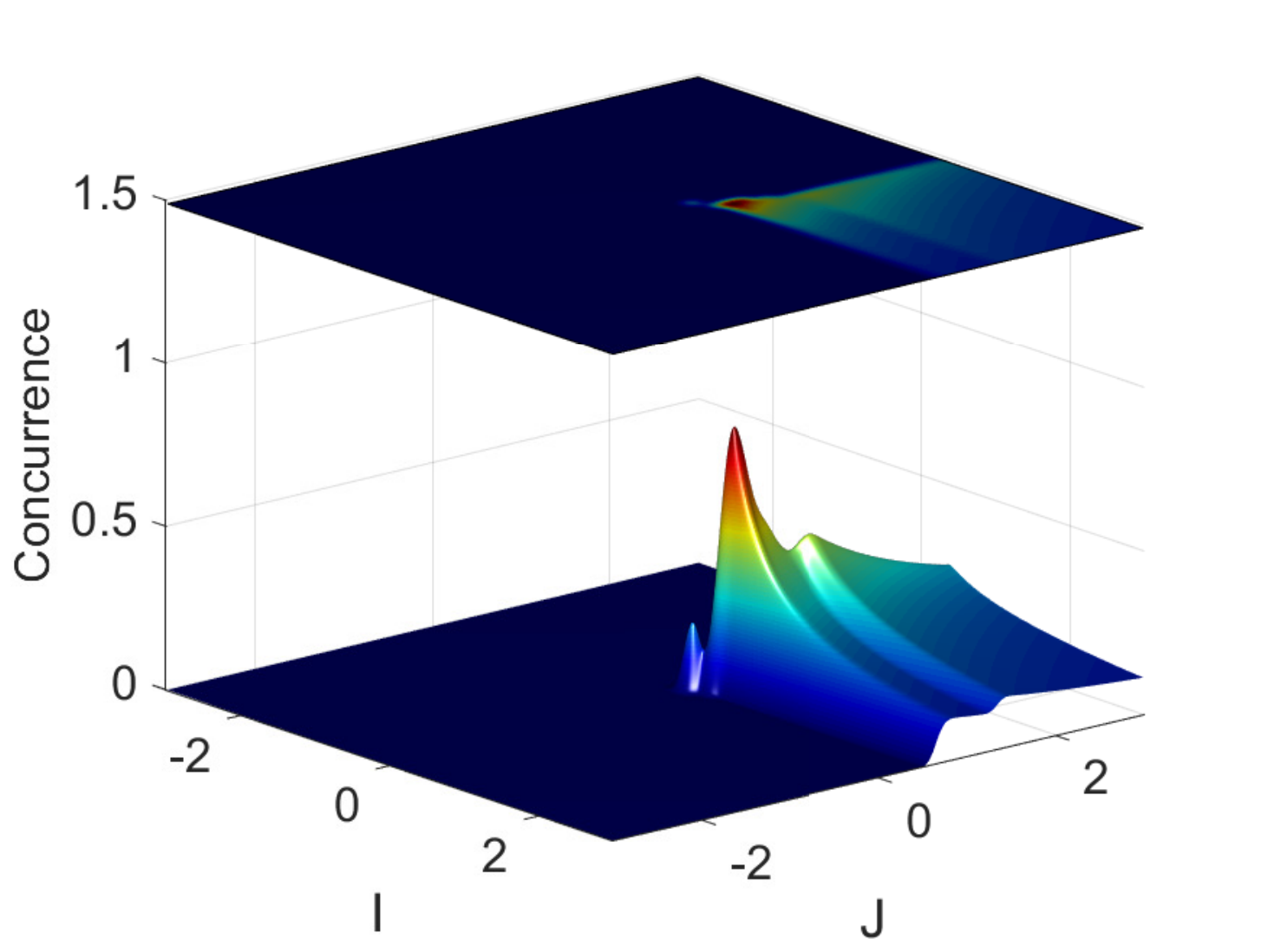}
(c) \includegraphics[width=0.43\columnwidth]{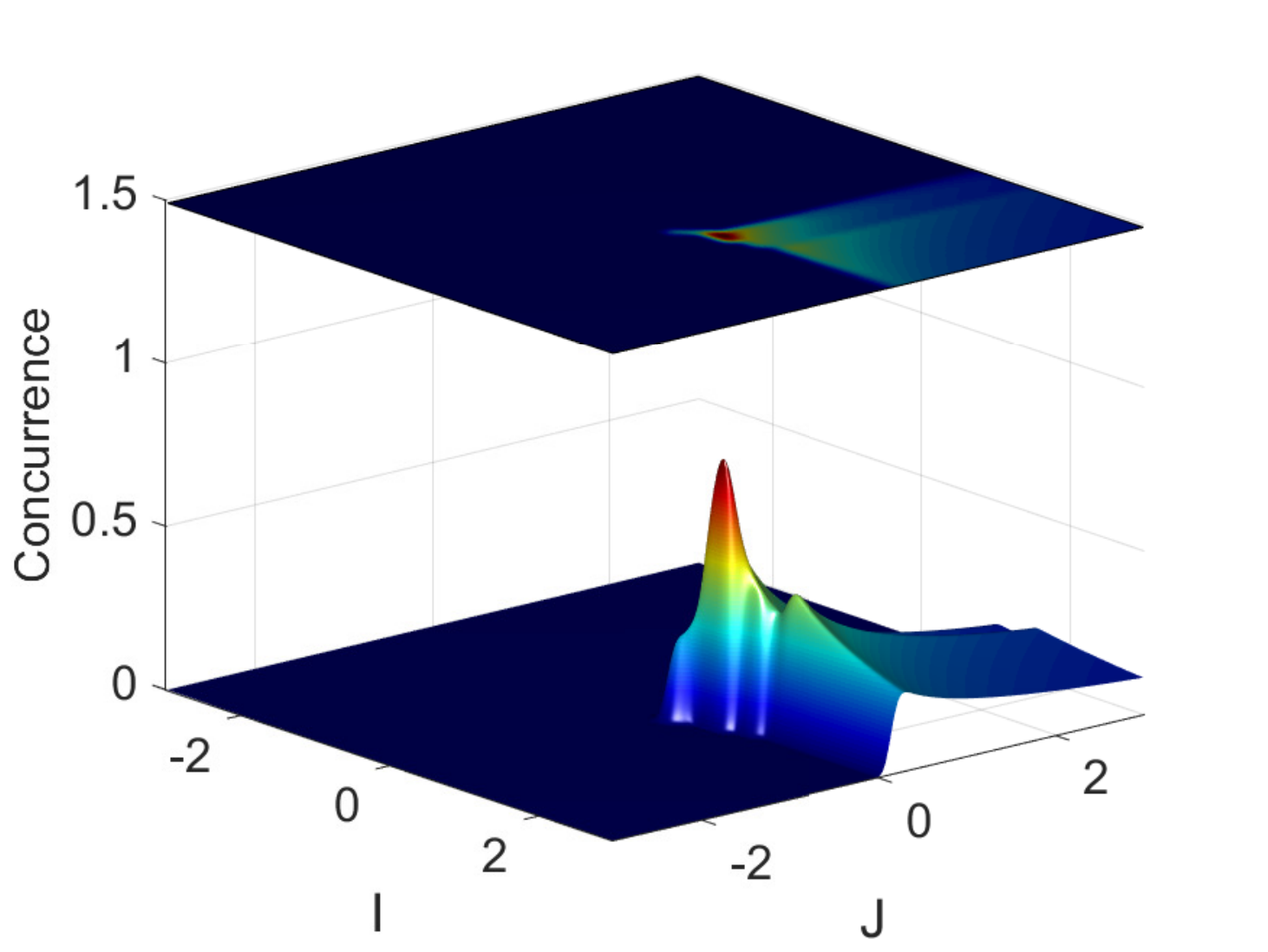}\\
(d) \includegraphics[width=0.43\columnwidth]{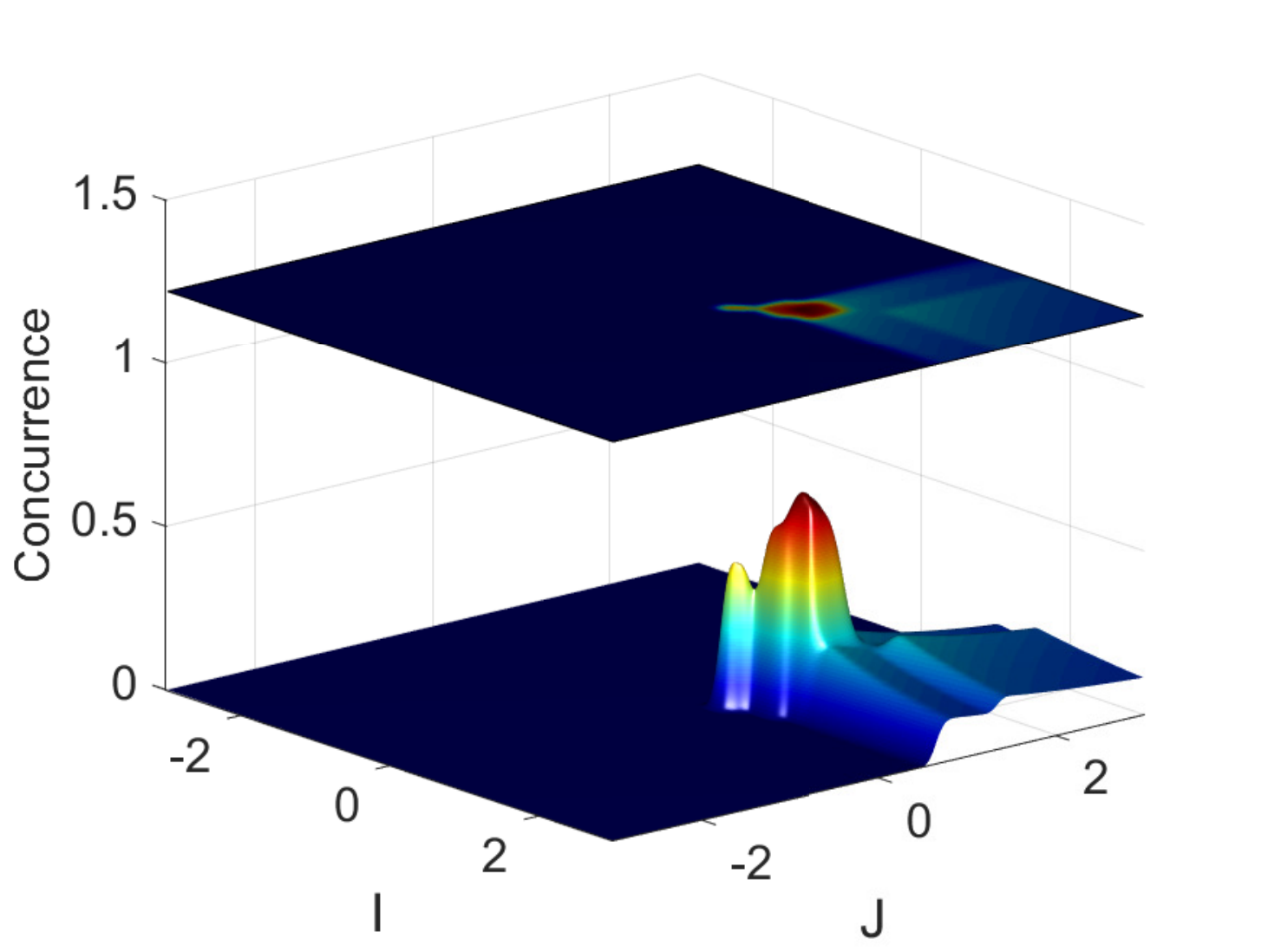}
(e) \includegraphics[width=0.43\columnwidth]{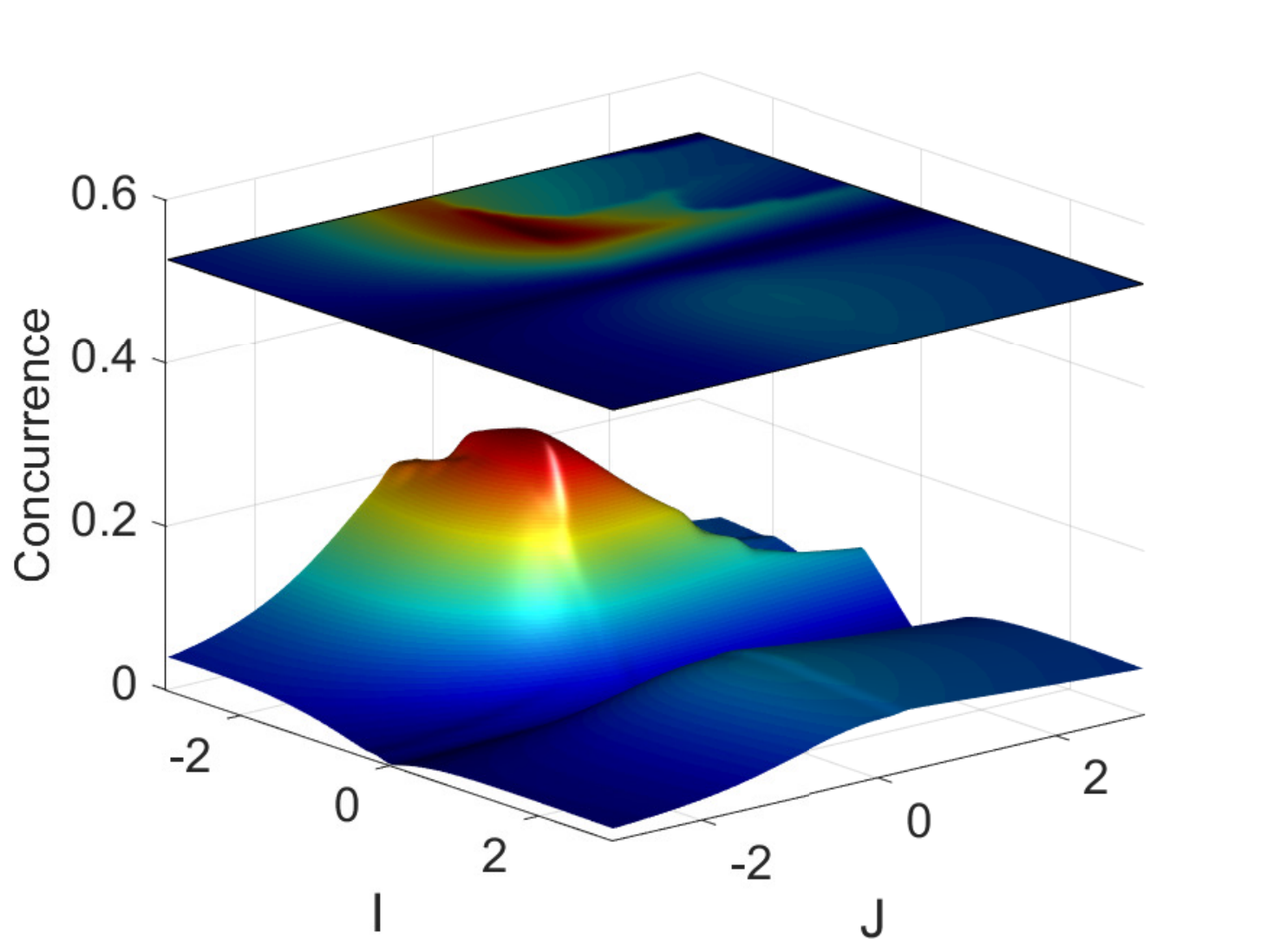}

\caption{Entanglement $C$\,(\ref{conc}) for intersubsystem exchange (\ref{eqKHTS2})
$ \sim \left( {\hat{\mathbf{S}}}_{\mathbf{i}} {\hat{\mathbf{S}}}_{\mathbf{j}}\right) \left(T_{\mathbf{i}}^{z}T_{\mathbf{j}}^{z}\right)$
with positive $K=+1$. At zero external fields, the entangled state is realized at a the $J, J >0$ domain.
(a) $\mathcal{H}_{s} = \mathcal{H}_{t} \ll 1$.
(b) $\mathcal{H}_{s} = 1, \mathcal{H}_{t} \ll 1$.
(c) $\mathcal{H}_{s} \ll 1, \mathcal{H}_{t} = 1$.
(d) $\mathcal{H}_{s} = 1$ and $\mathcal{H}_{t} = 1$.
(e) Staggered fields $|\mathcal{H}_{s}| = |\mathcal{H}_{t}| = 1$ in both subsystems.
Here, $\mathcal{H}_{s}$ and $\mathcal{H}_{t}$ stand for external fields in spin and pseudospin subsystems.
}
\label{K=+1_sszz}
\end{figure}

Thus, in this section, we propose a criterion
(purely empirical) for indicating the region of quantum entanglement in complex many-particle systems. It requires neither checking the Bell's inequalities nor calculating the full density matrix. The corresponding two-site correlator can be determined either numerically, but with much less waste of resources, or even analytically~\cite{Valiul19_JL}.

\section{\label{sec:alter} Other types of spin-pseudospin interactions}

Hereinafter, we consider other possible types of spin-pseudospin interaction, that up to now have not been investigated, at least, in the context of entanglement.

\subsection{\label{sec:sszz} Pseudospin anisotropic Interaction:
$\hat{\bf{H}}_{ts} = \sum \left( {\hat{\mathbf{S}}}_{\mathbf{i}}{\hat{\mathbf{S}}}_{\mathbf{j}}\right) \left( T_{\mathbf{i}}^{z}T_{\mathbf{j}}^{z}\right)$
}

In this subsection, we consider  what changes in the entanglement pattern entail a nontrivial, less symmetric spin-pseudospin interaction.
This refers to the Hamiltonian (\ref{eqKHH})--(\ref{eqKHS}) with the interaction between the subsystems (\ref{eqKHTS2}) ---
Heisenberg-type interaction
in $\hat{\bf{H}}_{ts}$ relating for spins and Ising-type one for pseudospins (note that this kind of interaction along with~(\ref{eqKHTS1}) was already proposed in the pioneering work on spin-orbital physics in compounds of transition-metal elements~\cite{Kugel82_SPU}).

The most dramatic changes occur in the case of negative intersubsystem exchange constant $K=-1$. Here, in addition to the region $J, I \gtrsim 0$, a whole new region of significant entanglement $C$\,(\ref{conc}) arises (see Fig.\,\ref{K=-1_sszz}a). More than a half of the investigated region is occupied by entangled state separated by the trivial line $I = 0$.

For the other sign of the intersubsystem exchange constant $K = + 1$, there are no qualitative changes in the entanglement structure in Fig.\,\ref{K=+1_sszz}a in comparison with the similar one in Fig.\,\ref{K=+1_sstt}a for symmetric intersubsystem interaction~(\ref{eqKHTS1}). Here, the entanglement area is qualitatively the same, a distinct ``shark tooth'' is formed near the origin. On the other hand, quantitative changes in the fine structure are rather significant.

The response of the spin-orbital system to a nonzero field for $K=-1$ differs qualitatively from the previous case, see Fig.\,\ref{K=-1_sstt}b-e. The picture does not change qualitatively, when the external field is nonzero in the spin subsystem. Only a shift is observed along the corresponding coordinate axis ($J$) (Fig.\,\ref{K=-1_sstt}b). On the contrary, the external field in the pseudospin subsystem destroys the entangled state in a quarter of the phase plane ($ J> 0 $, $ I <0 $)(Fig.\,\ref{K=-1_sstt}c). If there is a nonzero external field in both subsystems, a sharp peak of entanglement is formed near the origin, see Fig.\,\ref{K=-1_sstt}d. As in Sec.\,\ref{sec:fields}, the mutual orientation of the fields does not affect the structure of entanglement significantly.

Finally, the staggered field drastically changes the whole entanglement pattern. Qualitatively picture seems to be rotated from zero-field case  by $\pi/4$ counterclockwise. Significant entanglement appears at half plane (pseudospin subsystem exchange $I<0$).

At $K = + 1$, Figs.\,\ref{K=+1_sszz}a-e, the zero-field maximum entanglement is localized nearly at a single point, and when an external field is nonzero in one of subsystems, there is a tendency to isolate the area of maximum entanglement from the rest of the region with zero entanglement. This tendency is especially pronounced for $\mathcal{H}_{s} \ll 1, \mathcal{H}_{t} = 1 $, see Fig.\,\ref{K=+1_sszz}c. When there are two nonzero external fields, a sharp peak of entanglement is formed near the origin Fig.\,\ref{K=+1_sszz}c, which is insensitive to the mutual orientation of the fields.

The effect of the staggered field (``counterclockwise rotation''), significant entanglement for $I<0$) is similar to that for $K = -1$ except inessential details. We can't help mentioning that Fig.\,\ref{K=-1_sszz}e, and Fig.\,\ref{K=+1_sszz}e resemble some of arts of Zaha Hadit.

\subsection{\label{sec:zzzz} Spin and pseudospin anisotropic interaction:
$\hat{\bf{H}}_{ts} = \sum \left( S_{\mathbf{i}}^{z}S_{\mathbf{j}}^{z}\right) \left( T_{\mathbf{i}}^{z}T_{\mathbf{j}}^{z}\right)$
}

Here, we discuss the case when the intersubsystem interaction is even less symmetrical and has an Ising form in the parts
in $\hat{\bf{H}}_{ts}$, referring to both spins and pseudospins. This is the Hamiltonian (\ref{eqKHH}) - (\ref{eqKHS}) with the interaction between the subsystems (\ref{eqKHTS3}). This interaction is the Ashkin--Teller one~\cite{AshkinPR1943}, although model (\ref{eqKHH}), (\ref{eqKHS}), (\ref{eqKHTS3}) technically differs from the Ashkin--Teller model for which the exact solution exists~\cite{KugelFNT1980}.

The case of $K=-1$ is of a special interest here. The entanglement is realized here at three-quarters of the phase plane (two of the entanglement regions are, of course, symmetrical), and there are three sharp peaks near the origin, besides, all areas of entanglement are separated by lines $J, I = 0$. For the opposite sign of the spin-pseudospin exchange $K = + 1$, the entanglement pattern is realized, which is qualitatively similar to Fig.~\ref{K=+1_sszz}a --- entanglement in the quarter of the phase diagram and the ``shark tooth'' near the origin (we will not give the corresponding figures).

\begin{figure}[b]
(a) \includegraphics[width=0.60\columnwidth]{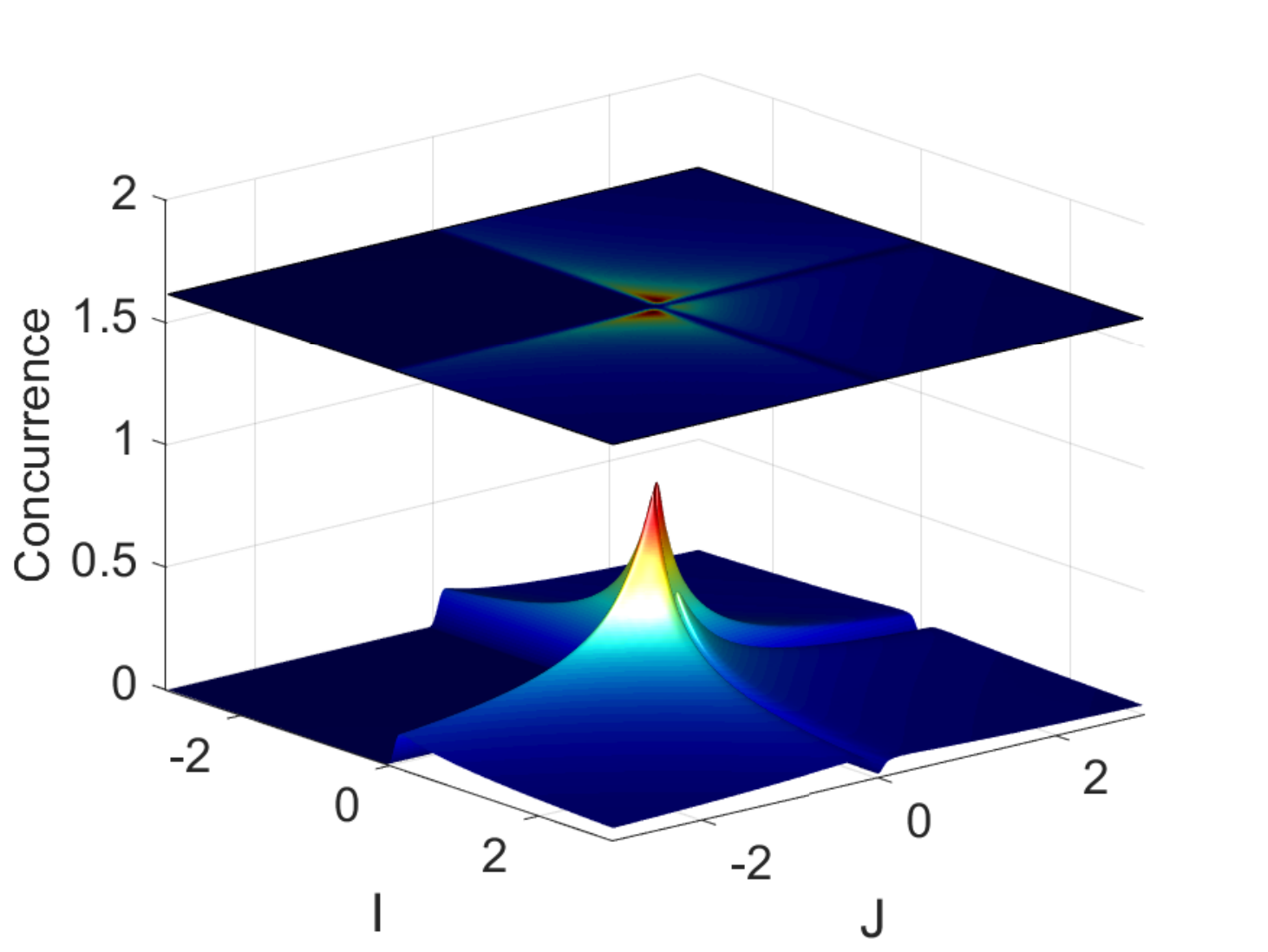}\\
(b) \includegraphics[width=0.43\columnwidth]{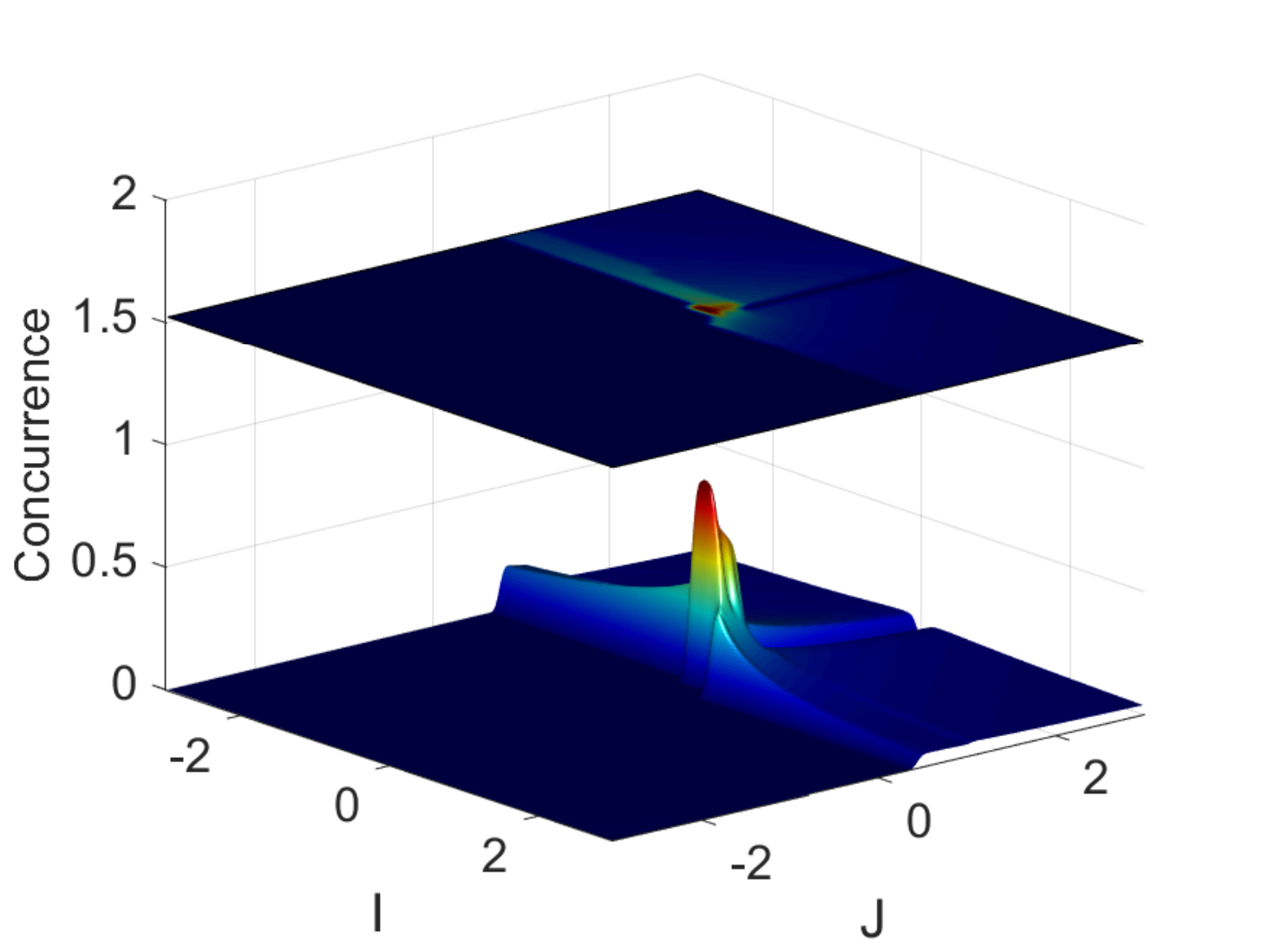}
(c) \includegraphics[width=0.43\columnwidth]{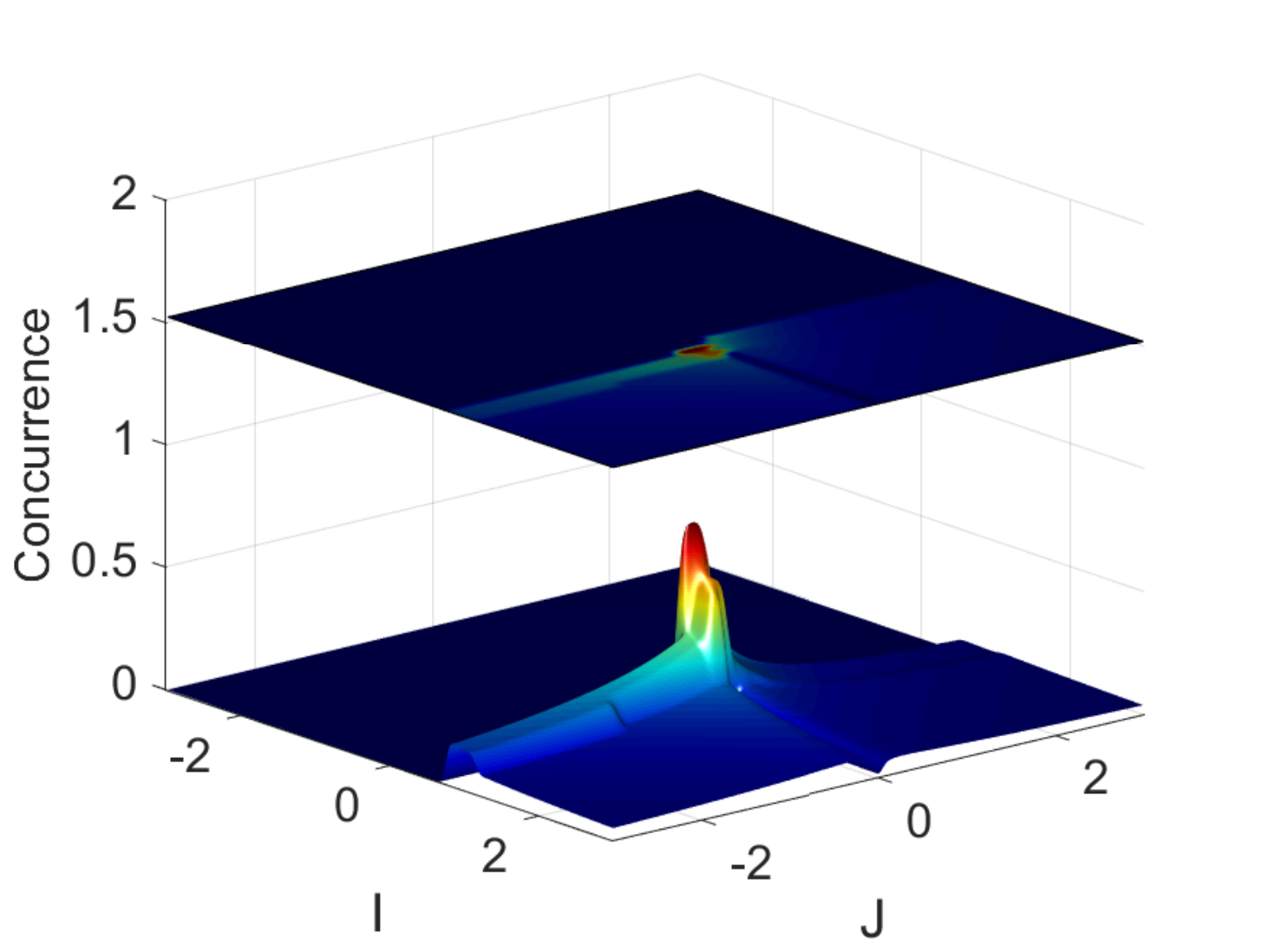}\\
(d) \includegraphics[width=0.43\columnwidth]{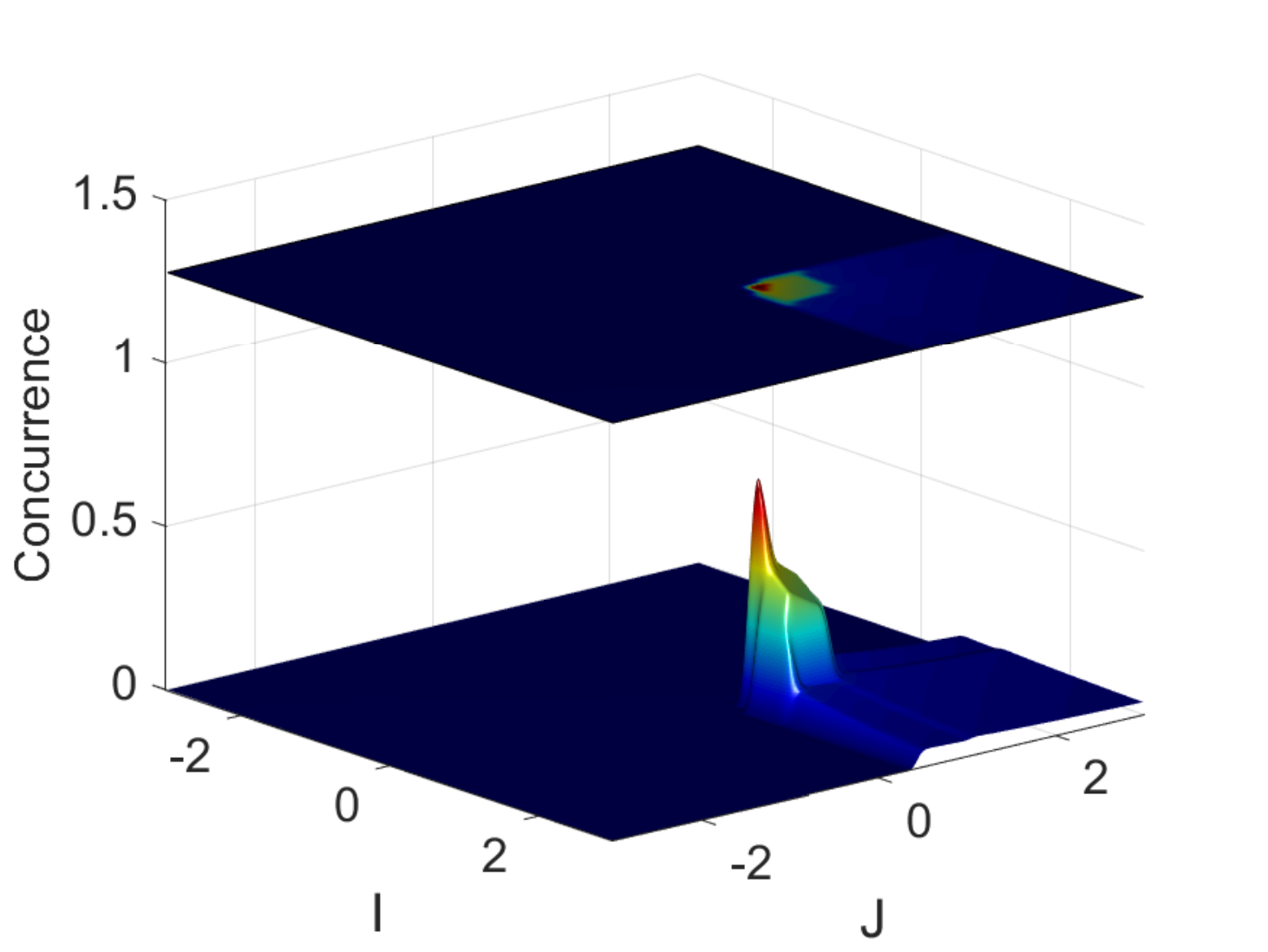}
(e) \includegraphics[width=0.43\columnwidth]{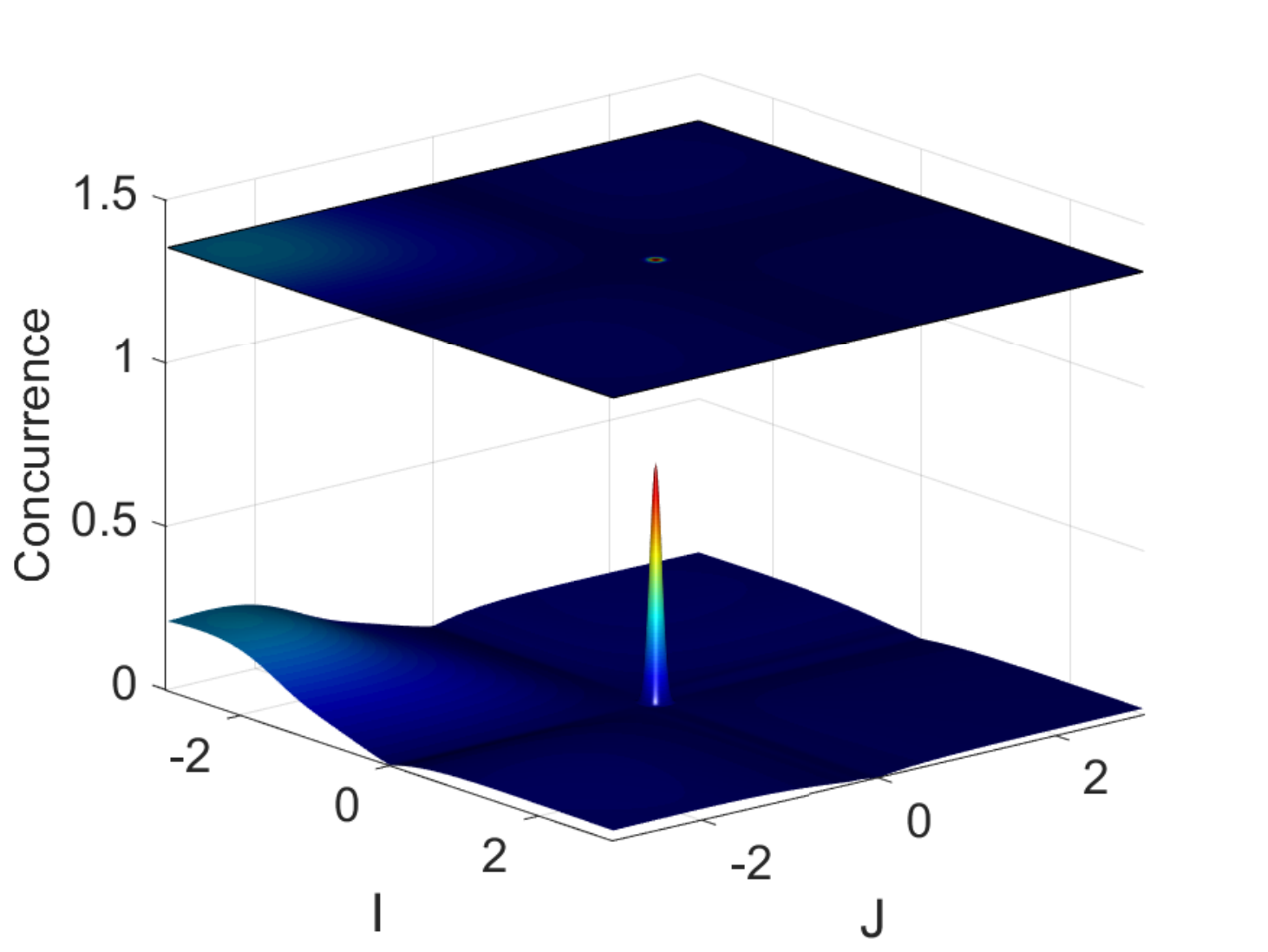}
\caption{Entanglement $C$\,(\ref{conc}) for intersubsystem exchange (\ref{eqKHTS3})
$\sim \left( S_{\mathbf{i}}^{z} S_{\mathbf{j}}^{z}\right)
\left( T_{\mathbf{i}}^{z}T_{\mathbf{j}}^{z}\right)$
with negative $K=-1$. At zero external fields, the entanglement is realized at 3/4 of the phase plane (except FM-FM region) with sharp maxima near the origin. When external field in spin/pseudospin subsystem is nonzero, the entanglement in the corresponding quadrant disappears completely.
(a) $\mathcal{H}_{s} = \mathcal{H}_{t} \ll 1$.
(b) $\mathcal{H}_{s} = 1, \mathcal{H}_{t} \ll 1$.
(c) $\mathcal{H}_{s} \ll 1, \mathcal{H}_{t} = 1$.
(d) $\mathcal{H}_{s} = 1$ and $\mathcal{H}_{t} = 1$.
(e) Staggered fields $|\mathcal{H}_{s}| = |\mathcal{H}_{t}| = 1$ in both subsystems.
Here, $\mathcal{H}_{s}$ and $\mathcal{H}_{t}$ stand for external fields in spin and pseudospin subsystems.
}
\label{K=-1_zzzz}
\end{figure}

Now, we address the nonzero field case. When the magnetic field is nonzero in one of the subsystems, the entanglement in the corresponding quadrant completely decays, and the situation in the other quadrants does not change qualitatively (with the increase of entanglement ``edges'' along one of the coordinate axes), see Figs.\,\ref{K=-1_zzzz}b--c. When two external fields are nonzero simultaneously, a sharp peak is formed near the coordinate origin with weak entanglement in the quadrant $ J> 0, I> 0$ and zero in the remaining regions, see Fig.\,\ref{K=-1_zzzz}d.

The effect of the staggered field is even more dramatic. The entanglement is almost or completely destroyed in the whole phase plane, except the peak at the origin. Note, however, that the concurrence is nonzero in the $J <0, I < 0$ quadrant.

The nonzero field case for the opposite sign $K = + 1$ differs in small details from the one just discussed, and we will not comment on it here.

\begin{figure}[t]
(a) \includegraphics[width=0.60\columnwidth]{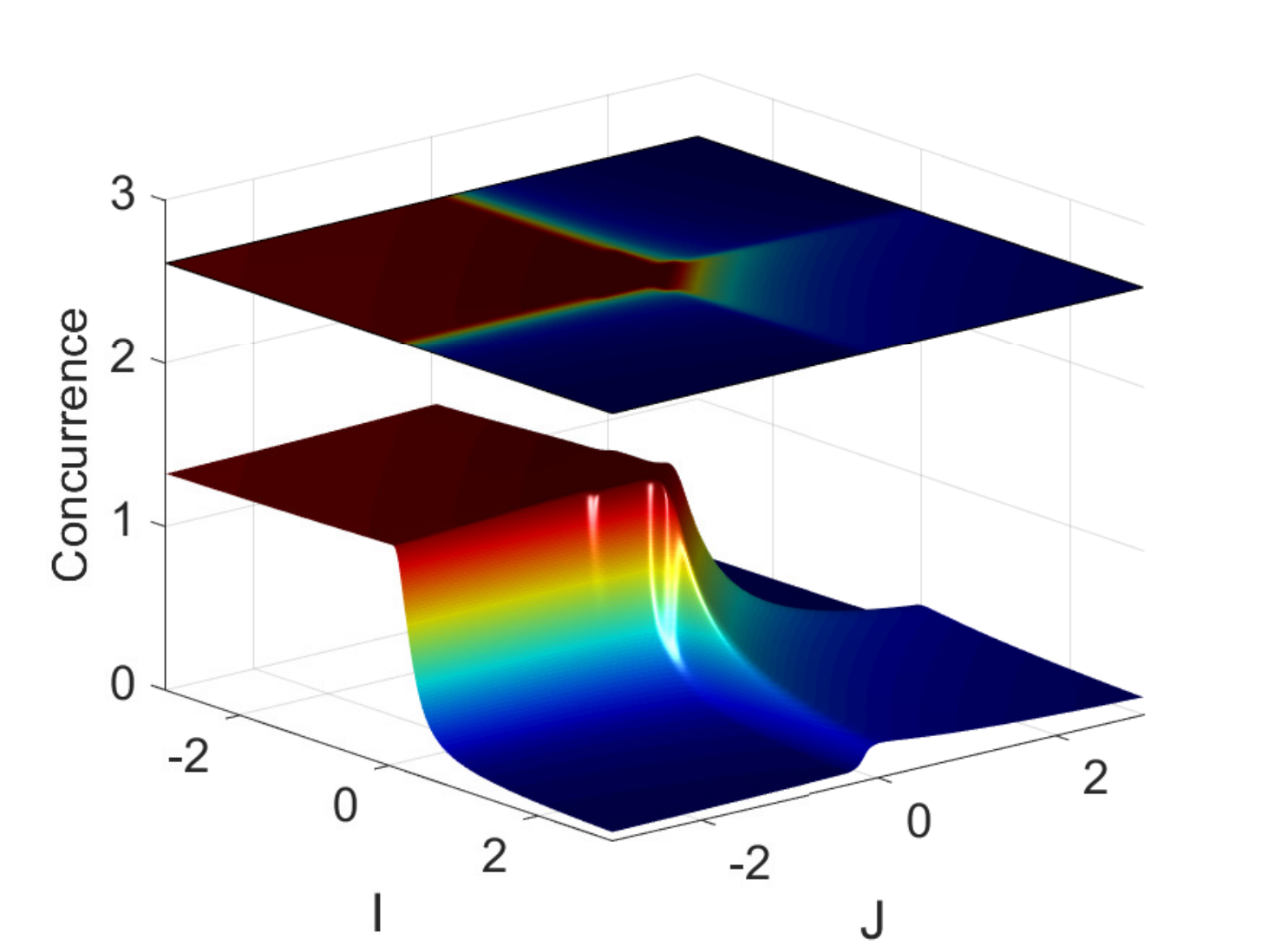}\\
(b) \includegraphics[width=0.43\columnwidth]{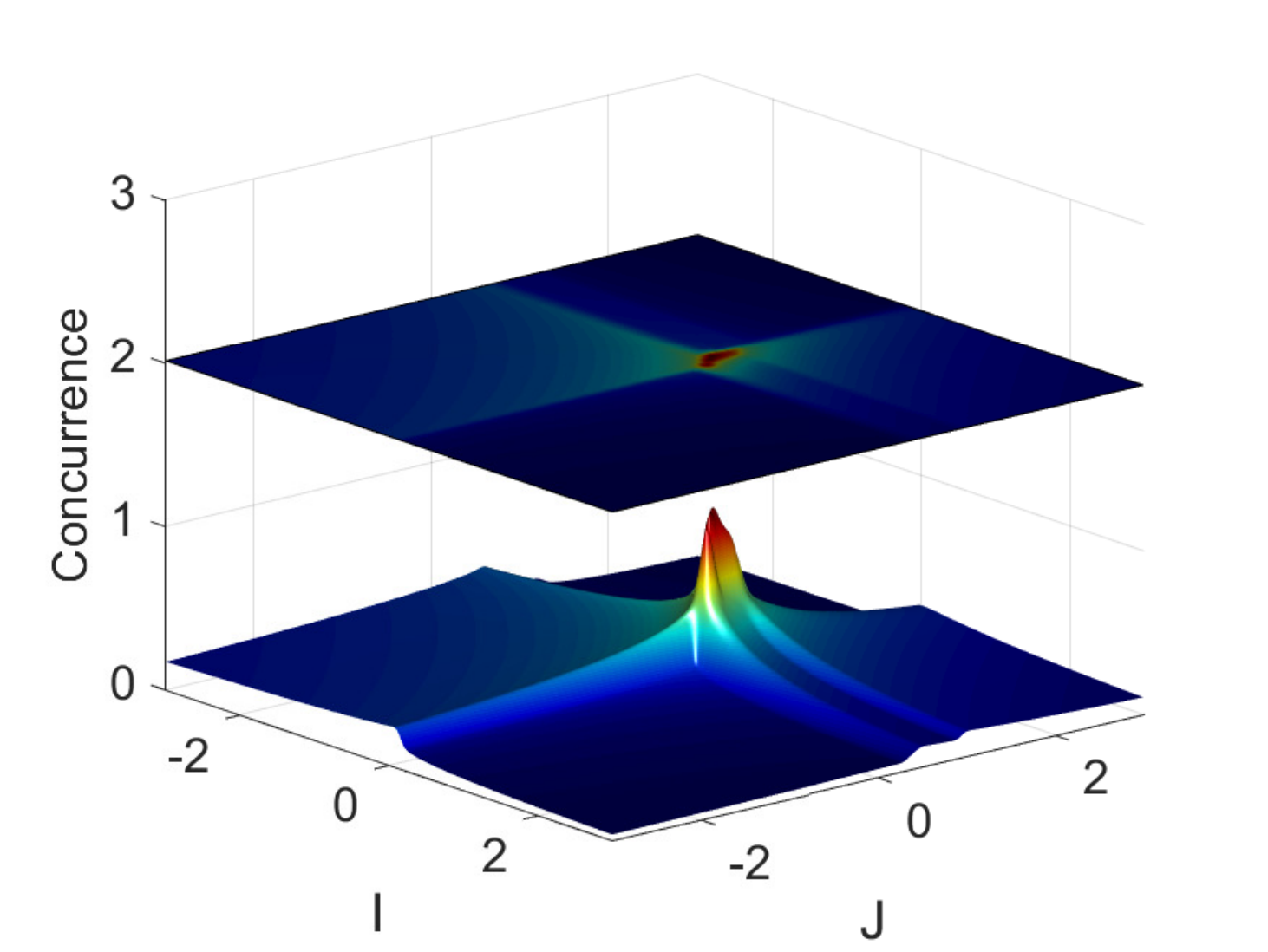}
(c) \includegraphics[width=0.43\columnwidth]{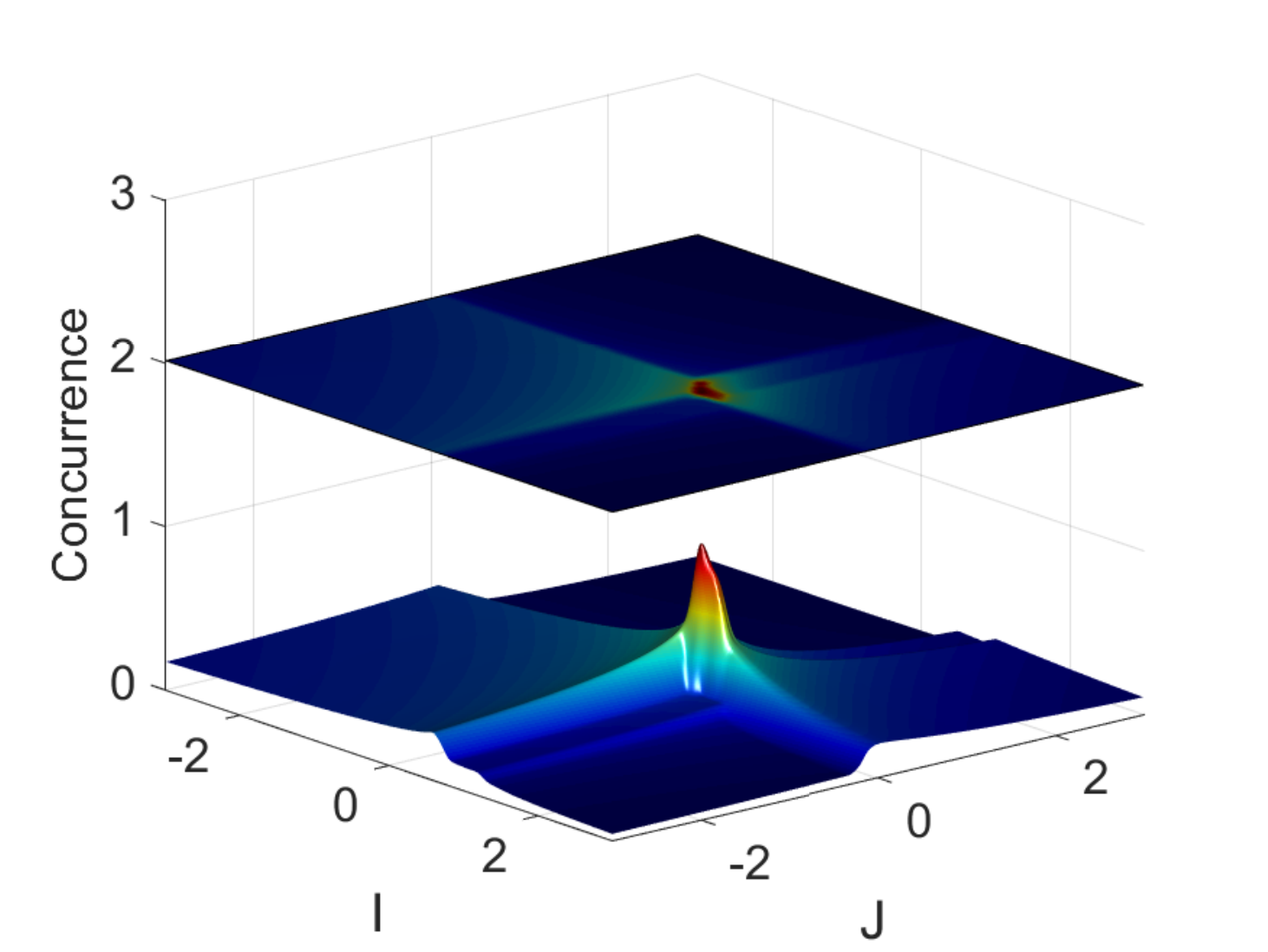}\\
(d) \includegraphics[width=0.43\columnwidth]{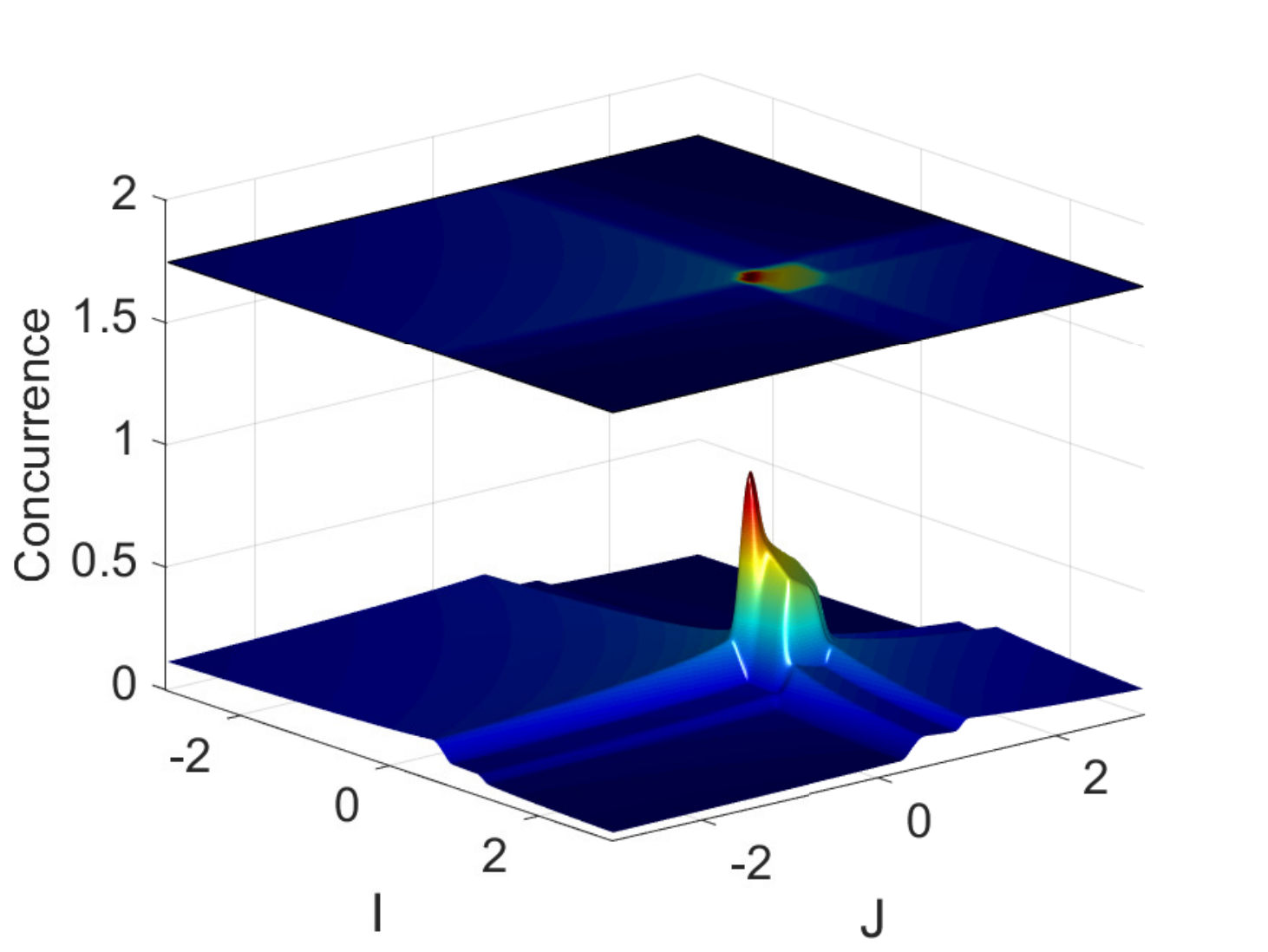}
(e) \includegraphics[width=0.43\columnwidth]{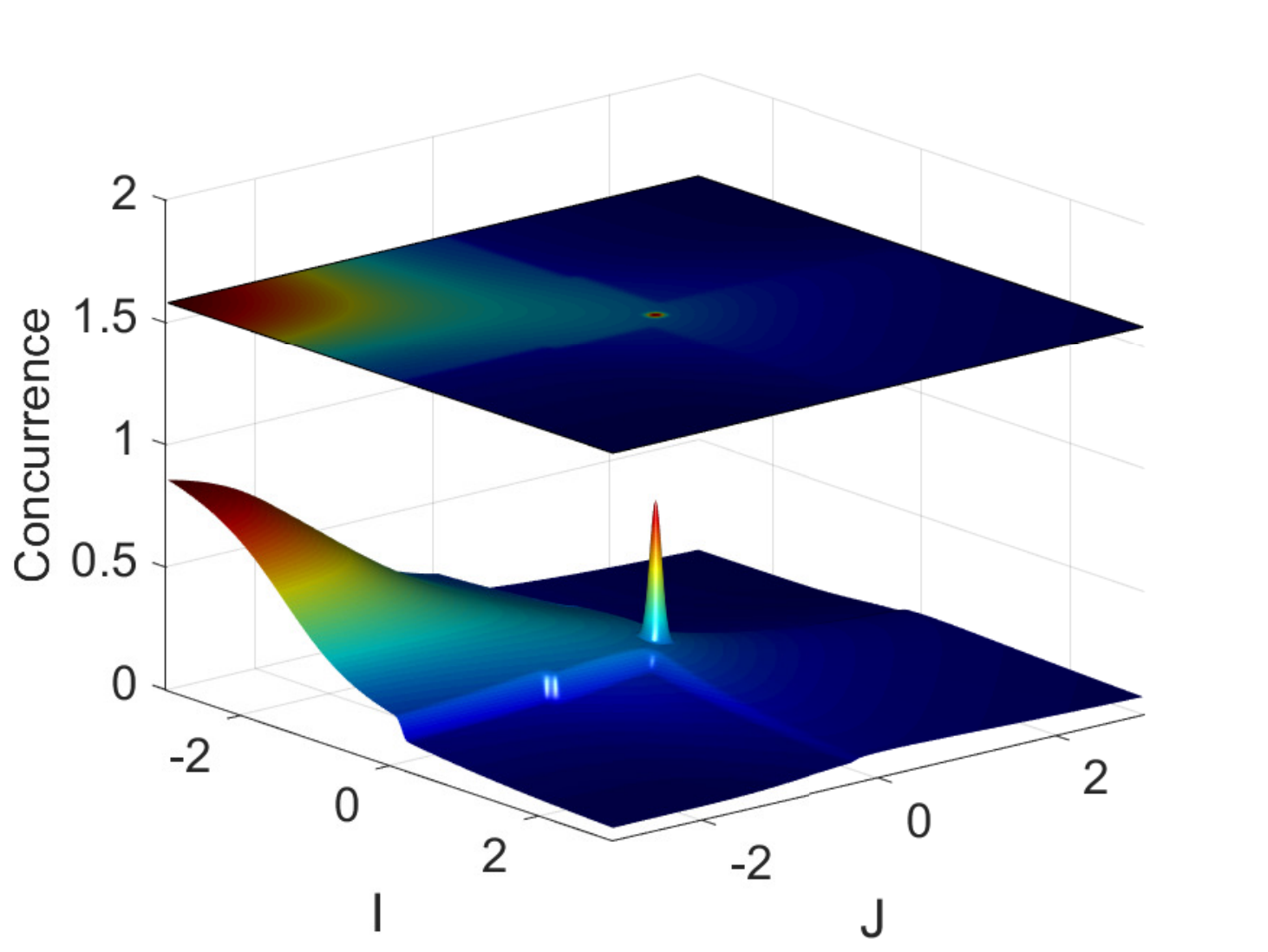}
\caption{Entanglement $C$\,(\ref{conc}) for intersubsystem exchange (\ref{eqKHTS4})
$\sim \left( S_{\mathbf{i}}^{\alpha}S_{\mathbf{j}}^{\alpha} T_{\mathbf{i}}^{\alpha}T_{\mathbf{j}}^{\alpha}\right))$
with negative $K=-1$. At zero external fields, the super-entanglement is formed in the FM-FM region of the phase plane. When external fields are nonzero, the entanglement disappears, leaving a sharp peak near the origin.
(a) $\mathcal{H}_{s} = \mathcal{H}_{t} \ll 1$.
(b) $\mathcal{H}_{s} = 1, \mathcal{H}_{t} \ll 1$.
(c) $\mathcal{H}_{s} \ll 1, \mathcal{H}_{t} = 1$.
(d) $\mathcal{H}_{s} = 1$ and $\mathcal{H}_{t} = 1$.
(e) Staggered fields $|\mathcal{H}_{s}| = |\mathcal{H}_{t}| = 1$ in both subsystems.
Here, $\mathcal{H}_{s}$ and $\mathcal{H}_{t}$ stand for external fields in spin and pseudospin subsystems.
}
\label{K=-1_trian}
\end{figure}

\subsection{\label{sec:trian} Model interaction:
$\hat{\bf{H}}_{ts} = \sum \left( S_{\mathbf{i}}^{\alpha }S_{\mathbf{j}}^{\alpha }T_{\mathbf{i}}^{\alpha }T_{\mathbf{j}}^{\alpha }\right)$
}

Here, we consider even more exotic case: a model interaction (\ref{eqKHTS4})a. This interaction looks very peculiar (and slightly resembles the compass model~\cite{Brzezi07_PRB,Erikss09_PRB,Jackel09_PRL}), nevertheless we discuss it for completeness of classification. With both signs of $ K $, the most striking feature is the arising super-entanglement at $ J, I <0 $, that is, with the ferromagnetic both intersubsystem exchanges. The behavior of entanglement in the region $ J, I> 0 $ qualitatively (and semi-quantitatively) resembles the case of Fig.\,\ref{K=-1_sstt}a and Fig.\,\ref{K=+1_sstt}a.

Here, all the nonzero field cases are peculiar.
When a magnetic field is nonzero in any of the subsystems, the entanglement in the corresponding quadrant dramatically decays, leaving mainly a sharp peak near the origin, see Fig.\,\ref{K=-1_trian}b,c. The presence of magnetic fields in two subsystems results in a peak near the origin insensitive to the mutual direction of the fields (Fig.\,\ref{K=-1_trian}d).

The effect of the staggered field looks here like in the model just considered (compare Fig.\,\ref{K=-1_trian}e and Fig.\,\ref{K=-1_zzzz}e). The entanglement almost or completely absent in the whole phase plane, except the peak at the origin. The concurrence is considerable on within the $J <0, I < 0$ quadrant.

Since, as in the previous section, for $K=+1$, the concurrence structure appears to be qualitatively the same, we will not comment on this case.

\section*{Conclusions}

In this paper, the problem of quantum entanglement was addressed in terms of the behavior of finite chains described by different types of of two-spin models. The analysis was performed by the exact diagonalization technique allowing one to find out comprehensive quantitative information concerning the systems under study. We were mainly focused on the behavior of concurrence, which is a good numerical measure  of the entanglement. We determined the regions of pronounced entanglement at various relations  between the characteristic parameters of the models. We have also revealed certain similarities in the behavior of concurrence and that of the two-site correlation functions (the latter can be considered as a local indicator of entanglement).

We have also demonstrated the possibility to provide efficient control of the entanglement pattern by external fields (and by switching on nontrivial interactions). In particular, external fields can induce considerable entanglement in the areas, where zero-field entanglement is clearly absent. On the other hand, the inverse effect is possible --- the concerted action of the fields in both spin subsystems diminishing the entanglement.

We emphasize that due to the different physical origins of effective spin and pseudospin the applied fields  fields may be of a completely different nature, from the magnetic field to elastic stresses. For example, the simplest field-dependent part of a spin--orbital Hamiltonian has the form $hS^z +\Delta T^z$, where $h$ is the magnetic field in energy units and $\Delta$ is the energy gap induced by local distortions~\cite{StreltsovUFN2017}. Note that here the superscript $z$ correspond to the $z$ axis in different spaces, spin and orbital ones. Depending on the ground state of the main Hamiltonian, such fields can affect the ground state in various ways, thus either enhancing or suppressing the entanglement.

The common experimental realization the entanglement effects is related to the spin--orbital excitations, referred to as orbitons~\cite{Saitoh01_N,Brink01_PRL,Ishiha05_NJP,Tanaka04_NJP,Schmid07_PRB,Bencki08_NJP}. This issue has drawn recently and an additional interest in  connection to the so called Higgs and Goldstone modes in condensed matter physics~\cite{KovalevaJPCM2013,JuraschekPRB2020,MeierPRB2020}.

Ultracold atoms bring a new perspective to spin--orbital physics.
Namely, a broad class of Hamiltonians of this type can be simulated not only in the framwork of solid-state strongly correlated systems but also by ultracold atoms in the traps~\cite{Gross17_S,Belemu17_PRB,Belemu18_NJP}. Then, the Kugel--Khomskii model can also involve an integer spin.
In such experiments, a variety of artificial external fields can be introduced by tuning laser beams or by the trap geometry rearrangemnet.

However, the role of quantum entanglement in the spin--orbital (spin--pseudospin) excitations has not been addressed properly yet. We believe, that our present work could be a good step forward in this direction.

\acknowledgments
This work was supported by the Russian Foundation for Basic Research, project Nos. 19-02-00509, and 20-02-00015. K.I.K. acknowledges the support from the Russian Science Foundation, project No. 20-62-46047.
N.M.S. acknowledges the support from the Russian Science Foundation, project No. 18-12-00438.

The computations were carried out on MVS-10P at Joint Supercomputer Center of the Russian Academy of Sciences (JSCC RAS). This work has been carried out using also computing resources of the federal collective usage center Complex for Simulation and Data Processing for Mega-Science Facilities at NRC ``Kurchatov Institute'', http://ckp.nrcki.ru/.

%\bibliographystyle{apsrevlong_no_issn_url}
%\bibliography{main}

\end{document}